\documentclass[prl,aps,twocolumn]{revtex4}

\usepackage{amsfonts}
\usepackage{amssymb}
\usepackage{graphicx}
\usepackage{mathrsfs}
\usepackage{array}
\usepackage{amsmath}
\usepackage{lscape}
\usepackage{bbold}
\usepackage{ucs}
\usepackage{color}

\begin{document}

\title{Capacity-resolution trade-off in the optimal learning of multiple low-dimensional manifolds by attractor neural networks}

\author{Aldo Battista}
\author{R\'emi Monasson}

\affiliation{Laboratory of Physics of the Ecole Normale Sup\'erieure, CNRS UMR 8023 \& PSL Research, Paris, France}

\date{\today}

\begin{abstract}
Recurrent neural networks (RNN) are powerful tools to explain how attractors may emerge from noisy, high-dimensional dynamics. We study here how to learn the $\sim N^2$ pairwise interactions in a RNN with $N$ neurons to embed $L$ manifolds of dimension $D \ll N$.  We show that the capacity, {\em i.e.} the maximal ratio $L/N$,  decreases  as $|\log \epsilon|^{-D}$, where $\epsilon$ is the error on the position encoded by the neural activity along each manifold. Hence, RNN are flexible memory devices capable of storing a large number of manifolds at high spatial resolution. Our results rely on a combination of analytical tools from statistical mechanics and random matrix theory, extending Gardner's classical theory of learning to the case of patterns with strong spatial correlations.
\end{abstract}
\maketitle

%% Introduction %%

How sensory information is encoded and processed by neuronal circuits is a central question in computational neuroscience. In many brain areas, the activity of neurons, $\sigma$, is found to depend strongly on some continuous sensory correlate $\bf r$; examples include simple cells in the V1 area of the visual cortex coding for the orientation of a bar presented to the retina, and head direction cells in the subiculum or place cells in the hippocampus, whose activities depend, respectively, on the orientation of the head and the position of an animal in the physical space. Over the past decades, Continuous Attractor (CA) neural networks have emerged as an appealing concept to explain such findings, more precisely, how a large and noisy neural population can reliably encode `positions' in low-dimensional sensory manifolds, $\sigma = \Phi({\bf r})$, and continuously update their values over time according to input stimuli \cite{Amari77,tsodyks95,Benyshai95,Wong10,Zhong18}. 

\begin{figure}[ht]
	\centering
	\includegraphics[width=.483\textwidth]{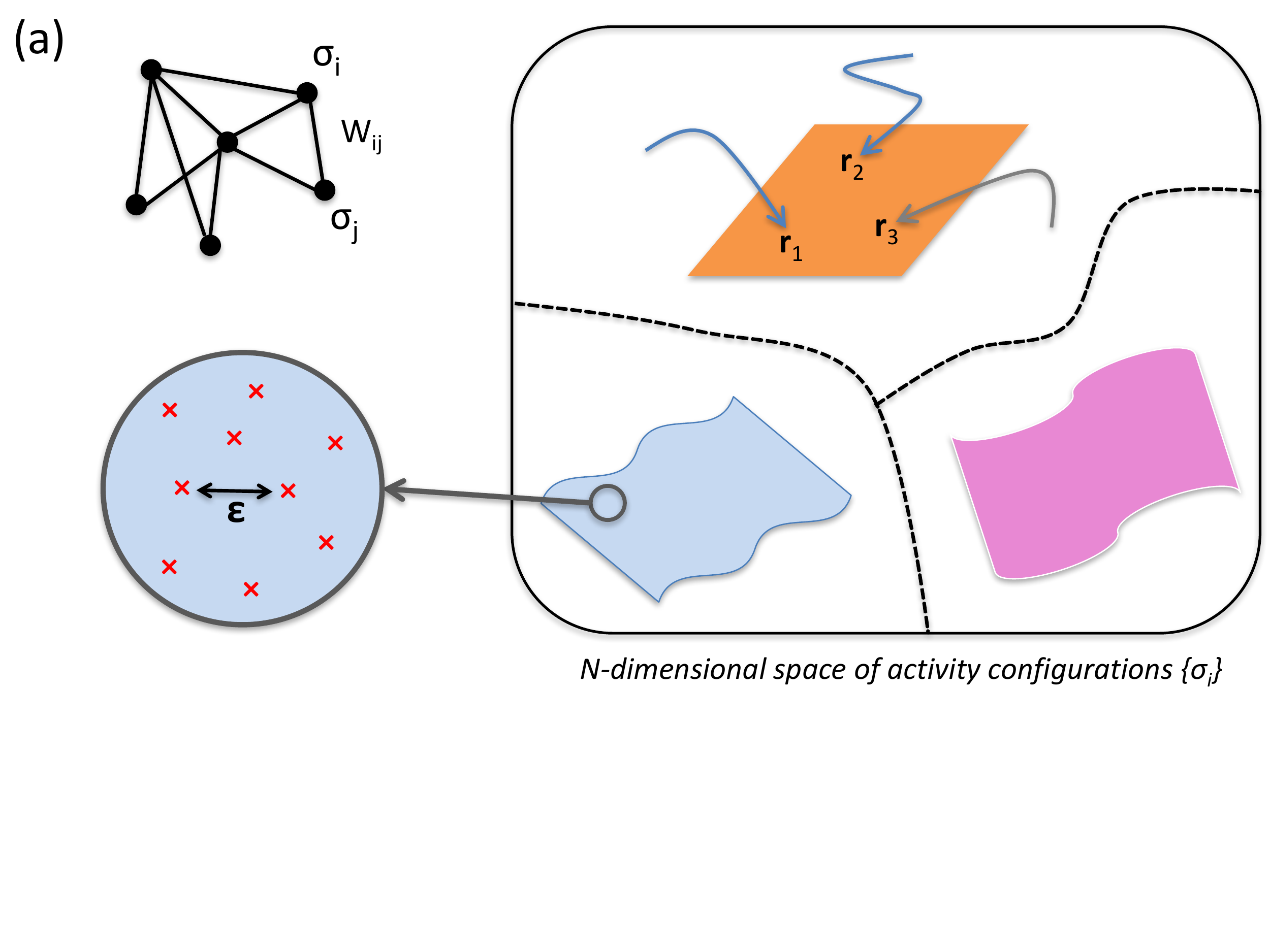}
	\vskip -1.3cm
	\includegraphics[width=.483\textwidth]{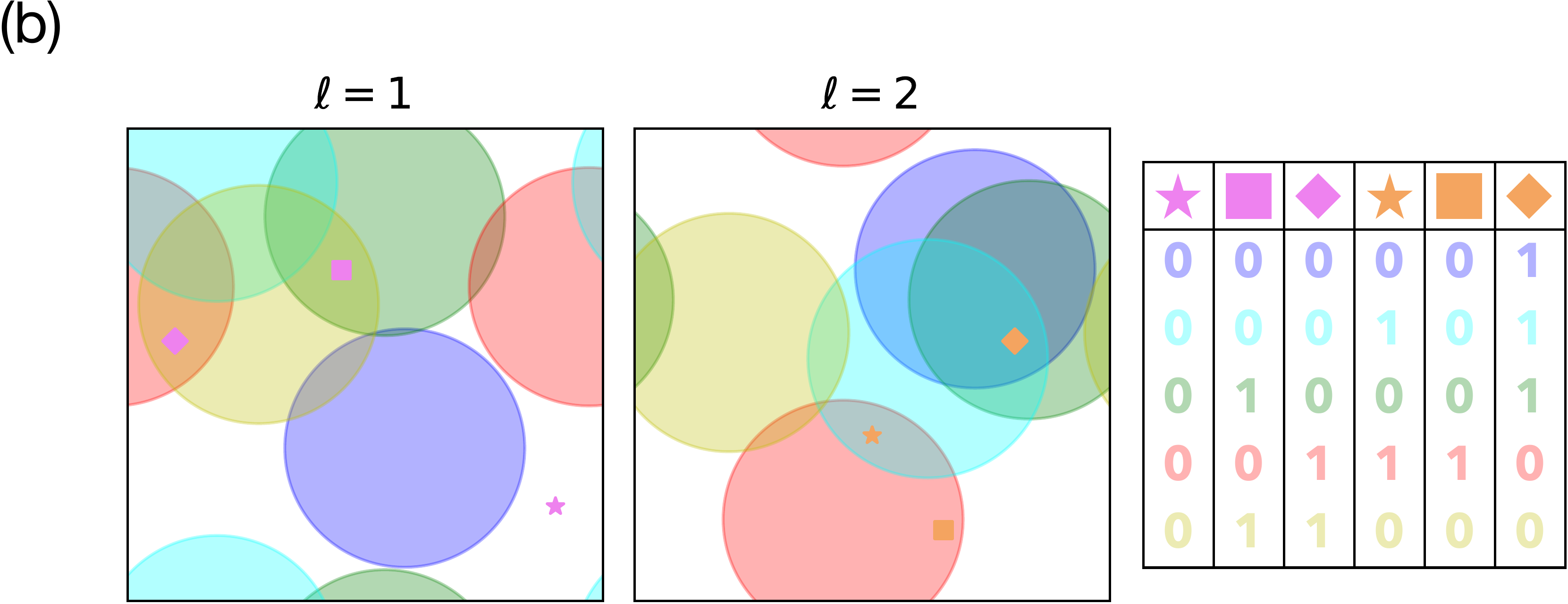}
	\caption{(a) A recurrent network with $N$ neurons and connectivity matrix ${\bf W}$ (top left) generates high-dimensional activity configurations attracted to multiple low-dimensional manifolds (right); on each manifold, we require to memorize $p$ points (bottom left, red crosses), whose separation defines the spatial resolution $\epsilon$. (b) Place fields (PF) of $N=5$ neurons in two maps, out of the three shown in panel (a). Each color identifies one neuron; the corresponding PF define the regions (with periodic boundary conditions) in the maps in which the neuron is active. The table lists, for each map, $p=3$ activity patterns corresponding to the marked points.}
        \label{fig:FIG_1}
\end{figure}

Models for the embedding of a CA in Recurrent Neural Network (RNN) generally assume that, after a Hebbian-like learning phase, the connection $W_{ij}$ between the neurons $i,j$ having their place fields centered in positions ${\bf r}_i$ and ${\bf r}_j$,  takes value
\begin{equation}\label{hebb1}
W_{ij} = w\big(|{\bf r}_{i}- {\bf r}_j| \big) \ ,
\end{equation}
where $|\cdot |$ denotes the distance in the sensory space. If $w$ is sufficiently excitatory at short distances and inhibitory at long ones, a bump state spontaneously emerges, in which active neurons tend to code for nearby positions in the sensory space. Weak external inputs suffice to move the bump and span the $D$-dimensional manifold of all possible positions $\bf r$ (Fig.~1(a)). This mechanism was  observed in the ellipsoid body of the fly, where a bump of activity points towards the heading direction \cite{Kim17}. Indirect evidences for the presence of CA have been reported, {\em e.g.} in the grid-cell system \cite{Yoon13} and in the prefontal cortex \cite{Wimmer14}. 

Hebbian connections (\ref{hebb1}) can be modified to embed in the same network of $N$ neurons multiple, unrelated CAs (Fig.~1(a)), such as multiple hippocampal spatial maps corresponding to different environments \cite{Alme14} or contextual situations \cite{Jezek11}. Assuming each one of the $L$ maps contributes equally to the learning process, connections take the form \cite{Samso97}
\begin{equation}\label{hebb2}
W_{ij} = \sum_{\ell=1}^{L} w\big(|{\bf r}_{i}^\ell- {\bf r}_j ^\ell | \big) \ ,
\end{equation}
where ${\bf r}_{i}^\ell$ is the center of the place field (PF) of neuron $i$ in environment $\ell$ (Fig.~1(b)). Theoretical calculations show that a bump state can exist (in any map) as long as $L<\alpha_c\, N$,  where $\alpha_c$ defines the critical capacity that can be sustained by the network \cite{Battaglia98,Monasson13}.

%% Problems with current theory %%

There are, however, serious practical and conceptual issues with the current theoretical understanding of multiple CAs based on (\ref{hebb2}). First, as soon as $L\ge2 $, the activity bump gets stuck in some preferred locations in the retrieved map due to the interferences coming from the other $L-1$ non-retrieved maps \cite{Cerasti13}. In other words, rule (\ref{hebb2}) does not define truly CAs, as large barriers oppose the motion of the bump along the map \cite{Monasson14}. The spatial error $\epsilon$ with which the environment is encoded, defined as the average discrepancy between any initial position ${\bf r}$ for the bump and the closest stable position in which it finally settles after neural relaxation dynamics, becomes quite large as $L$ increases (Fig. 2(a)). The issue of spatial resolution is also unclear from a theoretical point of view. Capacity calculations \cite{Battaglia98,Monasson13} require that a bump can form in any of the $L$ maps, in at least one position: they offer no guarantee about the existence of other memorized positions, and, more generally, about the value of $\epsilon$.

\begin{figure}[ht]
	\centering
	\includegraphics[width=.483\textwidth]{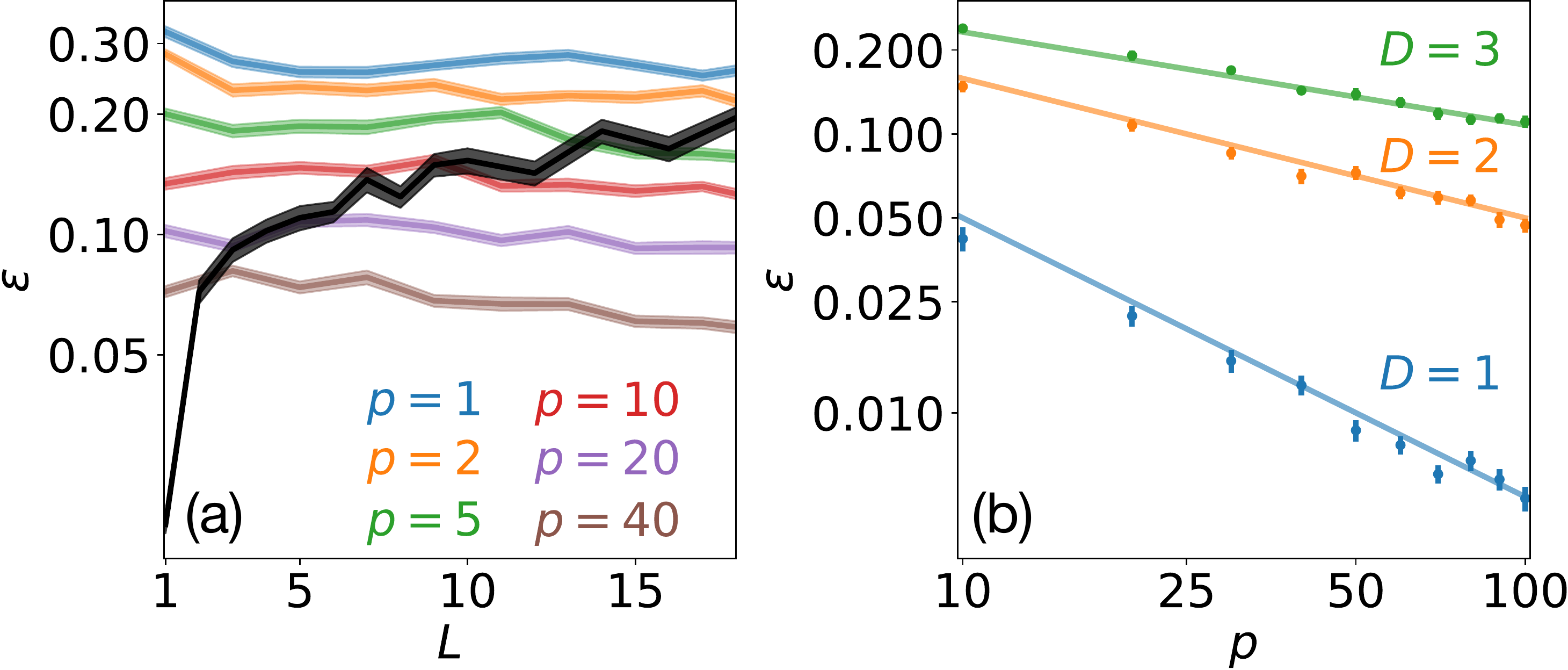}
	\caption{(a) Spatial error $\epsilon$ vs. number $L$ of two-dimensional maps in a network of $N=1000$ neurons. Black: rule (\ref{hebb2}), with $w(d)=e^{-d/.01}+w_0$, where $w_0<0$ enforces a fraction $\phi_0=.3$ of active cells. Colors: SVM results for different numbers $p$ of prescribed positions. Line widths show the error bars, see SM Sec.~I.E for details about the calculation of $\epsilon$. (b) Spatial error $\epsilon$ vs. number $p$ of positions in a network of $N=1000$ neurons storing $L=5$  maps, in dimensions $D=1,2,3$. Lines show the expected scalings $\epsilon \sim p^{-{1}/{D}}$ in log-log scale.}
        \label{fig:FIG_2}
\end{figure}

Secondly, the values of the critical capacity $\alpha_c$ with  rule (\ref{hebb2}) are generally quite low. It is reasonable to expect that the optimal storage capacity could be much higher:  a $\sim 15$-fold increase was found from the Hebb-rule critical capacity, $\simeq 0.14$ \cite{AGS85}, to the optimal capacity, $\alpha_c=2$ \cite{Gardner88} in the case of $0$-dimensional attractors, corresponding to the Hopfield model \cite{Hopfield82}. Optimal learning could also provide detailed insights on the statistical structure of the neural couplings $W_{ij}$, which could be compared to the physiological distribution of synaptic connections \cite{Brunel16}.

%% Overview

In this Letter, we present a theory of optimal storage of multiple quasi-continuous maps with prescribed spatial resolution in a RNN with $N$ binary neurons ($\sigma _i=0,1$) and real-valued, oriented connections $W_{ij}$. A map in this context is defined through the set of the input (place) fields of the $N$ neurons, each covering a volume fraction $\phi_0$ of the $D$-dimensional cube (Fig.~1(b)). In practice, the centers ${\bf \hat r}^{\ell}_i$ of the PFs are uniformly drawn at random in the cube, independently of each other, in all $\ell=1...L$ maps. For each map $\ell$, we draw uniformly at random $p$ positions ${\bf \hat r}^{\ell,\mu}$, $\mu=1...p$, and collect the $p$ corresponding patterns of activity: the neuron $i$ is active ($\sigma^{\ell,\mu}_i=1$) if the distance $|{\bf \hat r}^{\ell,\mu}-{\bf \hat r}^{\ell}_i|$ is smaller than the PF radius $r_c$, and silent ($\sigma^{\ell,\mu}_i=0$) otherwise (Figs.~1(a)\&(b)). 

\begin{figure}[ht]
	\centering
	\includegraphics[width=.45\textwidth]{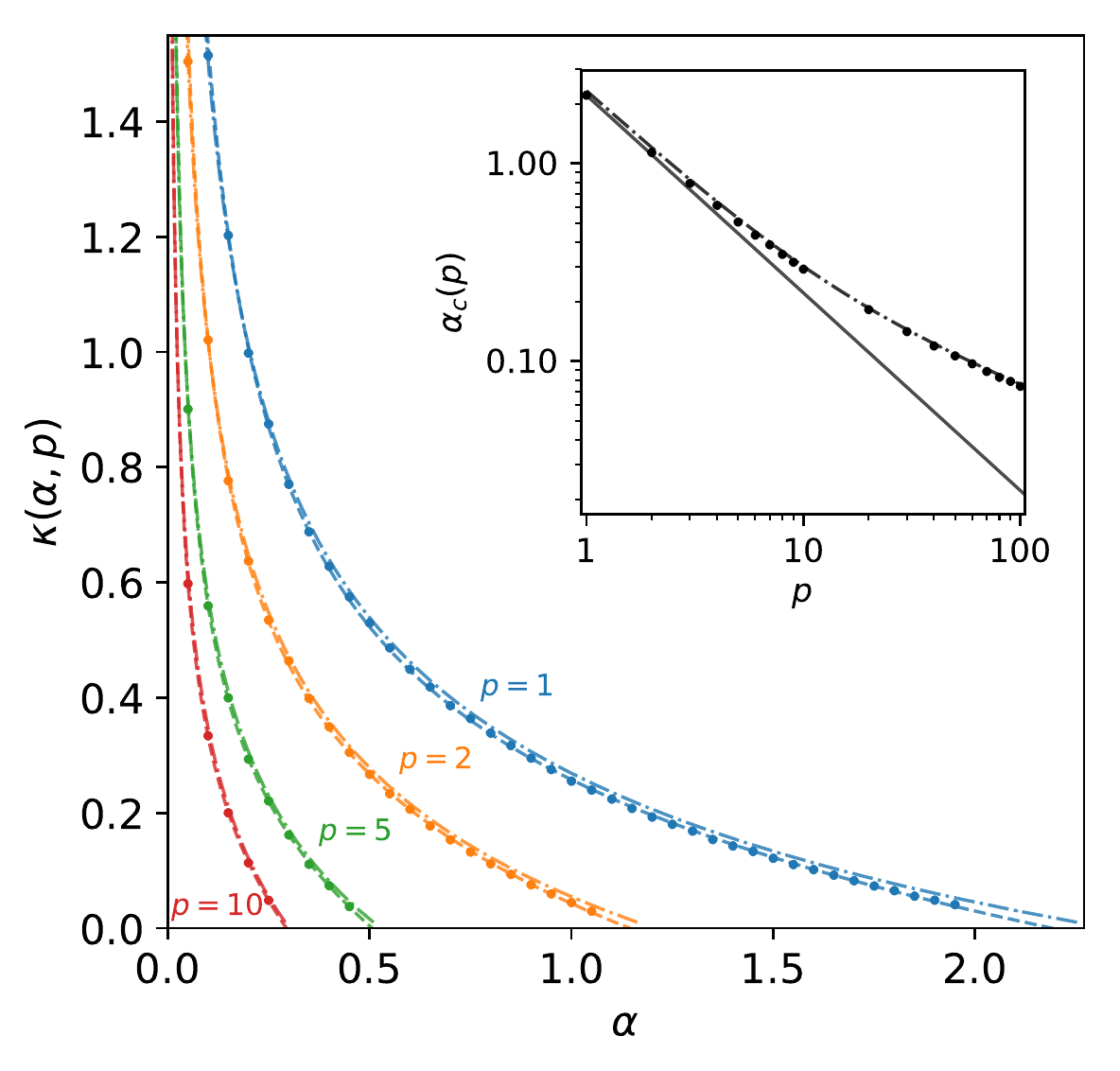}
\caption{Optimal stability $\kappa$ as a function of the load $\alpha$ and the number $p$ of positions. Dots: SVM results; Dashed lines: Gardner's theory  (\ref{alphac1}); Dashed-dotted lines: quenched PF theory (\ref{defga}). Parameter values: $D=2$, $\phi_0=.3$, $N=1000$ for SVM. 
Inset: $\alpha_c(p)$ decreases proportionally to $1/p$ (straight line) at low $p$, and much more slowly for large $p$. Dots indicate results from SVM ($N=5000$), averaged over $50$ samples, see SM Sec.~I.D for details on the estimation of $\alpha_c(p)$; the dot size indicates the maximal error bar. The dashed-dotted line shows the predictions from the quenched PF theory.}
\label{fig:FIG_3}
\end{figure}

%% Results with SVM

In order to learn these patterns we use Support Vector Machines (SVM) with linear kernels and hard margin classification (SM, Sec.~I.B) \cite{SVM} \cite{convex} \cite{sklearn}. We train $N$ SVM, one for every row $i$ in the coupling matrix $W_{ij}$, in which we consider the neuron $i$ as the output and the other $N-1$ neurons $j (\ne i)$ as the inputs \cite{Chung2018}. The training set $\{\sigma^{\ell,\mu}_i\}$ is common to all SVM. Once learning is complete, we normalize each row of the coupling matrix to $\sum _{j(\ne i)} W_{ij}^2=1$. SVM find the coupling matrix $\bf W$ maximizing the stability of the stored patterns,
\begin{equation}
\label{defkappa}
   \hskip -.1 cm \kappa= \max_{\bf W} \min_{\{i=1...N,\ell=1...L,\mu=1...p\}} \Big[  \big(2\sigma_{i}^{\ell,\mu}-1 \big)\sum_{j(\neq i)} W_{ij} \,\sigma_{j}^{\ell,\mu} \Big] \ .
\end{equation}

SVM couplings share some qualitative features with their Hebbian counterparts. First, the couplings $W_{ij}$ are correlated with the distances $d_{ij}^{\ell} = |{\bf r}_i^\ell - {\bf r}_j^\ell|$ between the PF centers of the neurons $i$ and $j$ in the different maps $\ell$, see SM Sec. I.C. Secondly, when simulating the trained network with simple rules for updating the neuron activities (SM, Sec.~I.E), the activity bump forms and diffuses within a map, and occasionally jumps to other maps \cite{Samso97,Monasson14,Monasson15}. However, with the maximal-stability learning rule, the spatial error $\epsilon$ can be tuned at will by varying $p$, see Fig.~2(a). For a fixed $p$, $\epsilon$ remains remarkably stable as the load increases until its critical value is reached. This is in sharp contradistinction with the Hebb rule case, for which $\epsilon$ quickly increases with the number of maps. The $p$ patterns form a discrete approximation of the map, with average spatial error scaling as $\epsilon = p^{-1/D}$, {\em i.e.} as the typical distance between neighboring points (Figs.~1(a)\&2(b)).  
 
The optimal stability $\kappa$ (\ref{defkappa}) is shown in  Fig.~3 as a function of the load $\alpha=L/N$ and of the number $p$ of prescribed fixed points; it is much higher than the maximal stability achievable with  rule (\ref{hebb2}) after optimization over the interaction kernel $w$, see SM, Sec.~I.D. As expected, $\kappa (\alpha, p)$ is a decreasing function of $\alpha$ and $p$: increasing the number of maps or enforcing finer spatial resolution reduces the stability. 
The value of the load at which $\kappa (\alpha, p)$ vanishes defines the critical capacity $\alpha_c(p)$, that is, the maximal load sustainable by the network as a function of the required spatial resolution.  Figure~3(inset) shows that $\alpha_c(p)$ decreases proportionally to $1/p$ at low $p$, and then much more slowly as $p$ grows. For small $p$, all $L\times p$ patterns are roughly independent, and we have $\alpha_c(p)\simeq \frac {\alpha_c(1)}p$, where $\alpha_c(1)$ is the capacity of the perceptron with independent, biased patterns having a fraction $\phi_0$ of active neurons \cite{Gardner88}. As $p$ gets large, substantial redundancies between the $p$ patterns within a map appear, as nearby positions define similar patterns (Fig.~1(b)), and the capacity is expected to decrease less quickly with $p$. The cross-over takes  place at $p_{c.o.}\sim 1/\phi_0$ (SM, Sec.~I.D). The non-trivial behavior of $\alpha_c(p)$ when $p \gg p_{c.o.}$ will be characterized in the theoretical study below.

%% Theory: Exact

Gardner's framework \cite{Gardner88} can, in principle, be applied to the optimal couplings corresponding to maximal stability $\kappa$ (\ref{defkappa}). Following standard calculations (SM, Sec.~II.A \cite{spinglass}), we find that the maximal load at fixed $\kappa$ and $p$ is given by
\begin{equation} \label{alphac1}
\alpha_c (\kappa,p) = 1/ \min _m \; \langle E_p ({\cal \hat R},{\cal Z},m;\kappa) \rangle_{{\cal \hat R},{\cal Z}}  \ ,
\end{equation}
where the minimum is taken over $m=\phi_0 \sum _{j(\ne i)} W_{ij}$. In the formula above, $\langle\cdot \rangle$ denotes the average over the vectors ${\cal \hat R}=({\bf \hat r}_1,...,{\bf \hat r}_p)$ of $p$ positions ${\bf \hat r}_\mu$ drawn uniformly at random in the $D$-dimensional cube, and ${\cal Z}=(z_1,...z_p)$ drawn from the multivariate centered Gaussian distribution with ${\cal \hat R}$--dependent covariance matrix 
\begin{equation}\label{defgau}
\boldsymbol\Gamma_{\mu, \nu} ({\cal \hat R}) = \Gamma\big(|{\bf \hat r}_\mu - {\bf  \hat r}_{\nu}|\big) - \phi_0^2 \ .
\end{equation}
Here, $\Gamma (d)$ is the overlapping volume between two PFs, whose centers are at distance $d$ from one another; hence, $\Gamma(0)=\phi_0$. Function $E_p$ in (\ref{alphac1}) is defined through
\begin{widetext}
\begin{equation} \label{deff}
E_p({\cal \hat R},{\cal Z},m;\kappa) = \min _{\{t_\mu  \ge \kappa + m, \mu=1...p \}}  \sum _{\mu,\nu=1}^p \bigg( t_\mu-z_\mu-2\,m\,\Phi(|{\bf  \hat r}_\mu|)\bigg) \, \boldsymbol\Gamma^{-1}_{\mu,\nu} ({\cal \hat R}) \, \bigg( t_{\nu}-z_{\nu}-2\,m\,\Phi(|{\bf \hat  r}_{\nu}|)\bigg) \ ,
\end{equation}
\end{widetext}
where $\Phi (d)=1$ if $d<r_c$, 0 otherwise; $r_c$ is the radius of the PF, {\em i.e.} the smallest number such that $\Gamma(2\,r_c)=0$. 

In practice, computing $\alpha_c(\kappa,p)$ from (\ref{alphac1}) is quite involved from a numerical point of view, as it requires to solve the $p$-dimensional semi-definite quadratic optimization problem in (\ref{deff}), as well as to average over the random vectors $\cal \hat R$ and $\cal Z$. This can be accurately done for small enough $p$, with results in excellent agreement with the SVM simulations, see Fig.~3. Notice that, for $p=1$, our calculation reproduces Gardner's critical capacity $\alpha_c(1)$ for independent and biased patterns (SM, Sec.~II.B). This is expected as spatial correlations between patterns within a map appear when $p\ge 2$. 

%% Theory: Gaussian

Formula (\ref{alphac1}) seems, unfortunately, intractable for large $p$. The intricate dependence on $p$, {\em e.g.} showing up through the Gaussian correlations between the $p$ random fields $z_\mu$ in (\ref{deff}), stems from the average (in each map $\ell$) over the $N$ PF centers, $\{{\bf r}_i^\ell\}$, at fixed positions $\{{\bf r}^{\ell,\mu}\}$. To avoid introducing these correlations and have an explicit dependence on the parameter $p$, we consider an alternative calculation scheme, where the $p$ positions in each map are averaged out, while keeping the $L\times N$ centers quenched. To further simplify the calculation we neglect in the effective action all terms of order $\ge 3$ in the couplings $W_{ij}$ \cite{Monasson93}; this Gaussian approximation is expected to be exact in the large-$p$ limit. Details about the calculation can be found in SM, Sec.~II.C.  Within our quenched PF theory the optimal load $\alpha_c(\kappa,p)$  is the root of $F$ defined through
\begin{eqnarray}\label{defga}
F(\alpha; m,q,U,V,T) &=& V\bigg( q+U - \frac {m^2}{1- \frac 4{g(U)}+4U}\bigg)   \\
&& \hskip -3.63 cm + T\bigg(1+ \frac{U\, g(U)-1}V\bigg)  - \alpha p (q-m^2) \int_{x}^\infty dz\,\frac{ e^{-\frac{z^2}2}}{\sqrt {2\pi}} (z-x)^2 \nonumber
\end{eqnarray}
with $x=\frac{m-\kappa}{\sqrt{q-m^2}}$. In (\ref{defga}), $m= \sum _{j(\ne i)} (2\,\boldsymbol {\cal C}_{ij}-\phi_0)W_{ij}$, $q=\sum _{j,k(\ne i)} W_{ij} \boldsymbol {\cal C}_{jk} W_{ik}$, and the Lagrange multipliers $U,V,T$ enforcing, respectively, the normalization of $W$ and the definition of the order parameters, are all chosen to optimize $F$. $\boldsymbol {\cal C}$ denotes the $N\times N$ multi-space Euclidean Random Matrix (ERM) 
\begin{equation}\label{matrixC}
\boldsymbol {\cal C}_{jk} \big( \{{\bf r}_i^\ell\}\big)= \frac 1L \sum_{\ell=1}^L \Gamma \big( | {\bf r}_j^\ell - {\bf r}_k^\ell | \big) \ , 
\end{equation}
with resolvent $g(U)= \frac 1N \text{Trace}\,(U\, {\bf Id}+\boldsymbol{\cal C})^{-1}$.
While ERM have been intensively studied in the literature \cite{ERM}, superimpositions of ERM mixing up different spaces have not been considered so far to our knowledge. The resolvent $g(U)$ can nevertheless be computed using tools from Random Matrix Theory \cite{vivo} \cite{Battista19}, and shown to be solution of the implicit equation
\begin{equation}
U =  \frac 1{g(U)} - \sum _{{\bf k} \ne {\bf 0} } \frac{ \alpha \, \hat \Gamma ({\bf k})}{\alpha  +g(U) \, \hat \Gamma ({\bf k})}  \ ,
\end{equation}
where the $\hat \Gamma ({\bf k})$'s are the components of the Fourier transform of $\Gamma$ on the $D$-dimensional infinite reciprocal cube. 

Resolution of these equations gives access to $\kappa (\alpha,p)$, in very good agreement with the numerical results obtained with SVM (Fig.~3). Small deviations can, however, be noticed and diminish with increasing $p$ as expected. The order parameters $q$ and $m$ are shown as functions of $p$ in Fig.~4, in good agreement with SVM results for large $p\ (\gg p_{c.o.})$.
The value of $p$ at which the confluence between the results from the quenched theory and SVM takes place is a decreasing function of the PF size  $\phi_0$ (SM, Sec.~II.D \cite{mizuseki} \cite{hussaini}) and of the map dimension $D$ (SM, Sec.~I.D). 

Due to the explicit dependence of $F$ on $p$ in (\ref{defga})  the asymptotic behaviour of the critical capacity can be analytically determined in the large--$p$ limit:
\begin{equation}\label{alphac2}
    \alpha_c (p) \sim A(D)\;\frac{\phi_0^{-(D-1)}} {(\log p)^{D} } \qquad (p\to\infty)\ ,
\end{equation}
where the constant $A$ is made explicit in SM, Sec.~II.C. Equation (\ref{alphac2}) is our main result. Informally speaking, the very slow decay of the critical capacity with $p$ (Fig.~3, inset) means that recurrent neural nets can efficiently store multiple spatial maps, even at high spatial resolution. More precisely, enforcing a strong reduction of the spatial error, such as $\epsilon \to \epsilon ^2$, results in a moderate drop of the maximal sustainable load, $\alpha_c\to \alpha_c/2^D$. In addition, the capacity is predicted to be a decreasing function of the PF size in  dimensions $D=2,3$, but not in dimension $D=1$. This asymptotic statement is qualitatively corroborated by SVM results, even for moderate values of $p$ (SM, Sec.~I.D). 

\begin{figure}[ht]
	\centering
	\includegraphics[width=.483\textwidth]{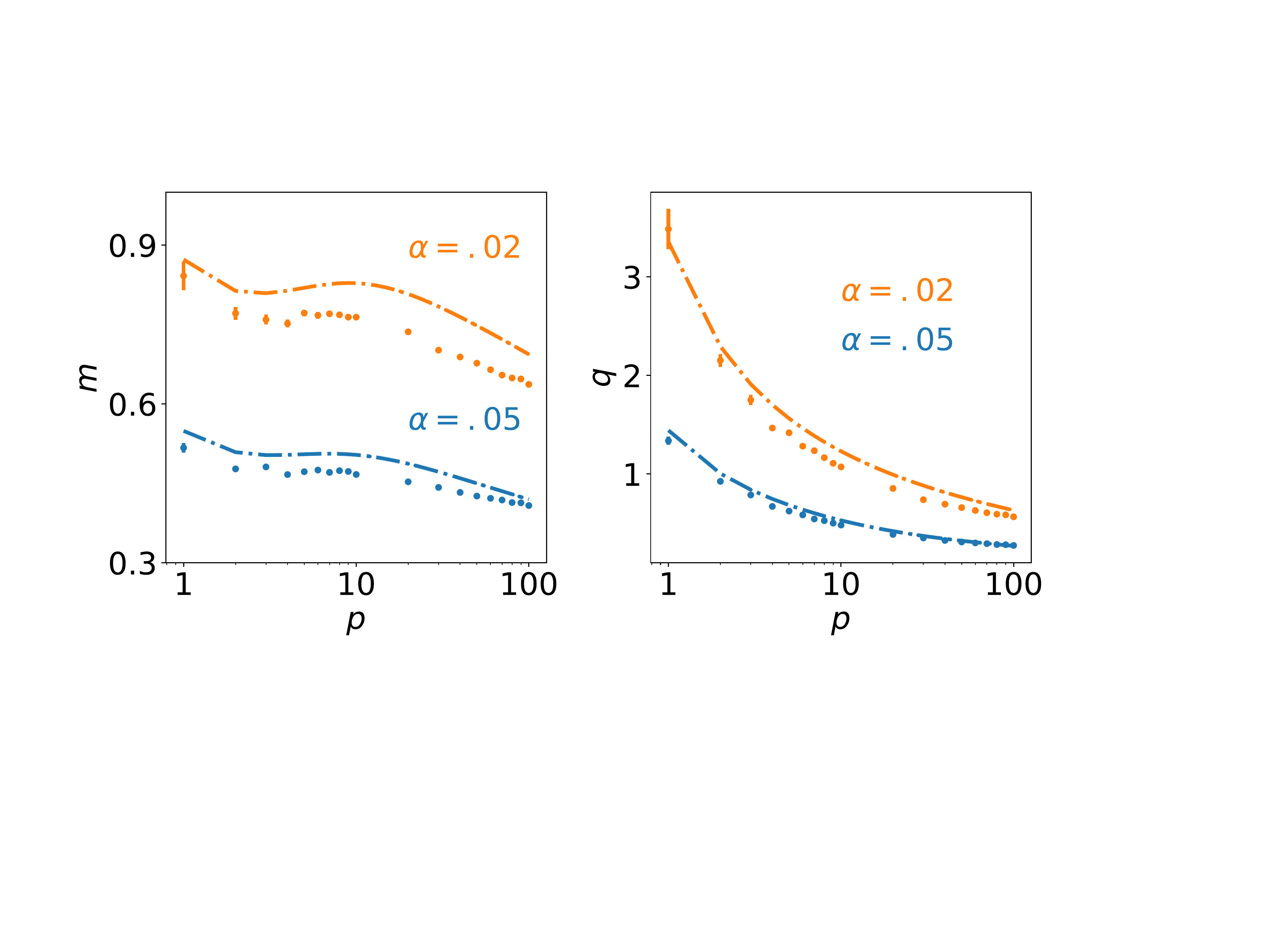}
\caption{Order parameters $m$ (left) and $q$ (right) vs. $p$. Dots: SVM results ($N=2500$), averaged over $50$ samples; Dashed-dotted lines: quenched PF theory (\ref{defga}). Parameters: $D=2$, $\phi_0=.3$, $\alpha=.02$ (top) and $.05$ (bottom), for which up to, respectively, $p_c\simeq 2500$ and $p_c\simeq 250$ points can be memorized. }
 \label{fig:FIG_4}
\end{figure}

%% Discussion

Many extensions of  the current work can be contemplated. First, our theory can be easily generalized to the case of spatial resolutions varying from map to map, by substituting $p$ with its average value over the maps in (\ref{alphac2}). This suggests that the fraction of maps with finest spatial resolution $\epsilon$ should not exceed $\sim \epsilon ^D$ when $\epsilon \to 0$, in order not to affect too much the critical capacity. 

Secondly, it would be very interesting to understand how much the scaling of $\alpha_c$ in (\ref{alphac2}) is robust against the choice of the parametrization $\Phi({\bf r})$ of the manifolds. We have shown that reducing the number of active neurons in each map and allowing for variations in the sizes of the PFs from neuron to neuron do not affect this scaling \cite{Battista19}.
While we have assumed here for the sake of simplicity that the distribution of points was statistically uniform across space, this need not be the case in practice. Experiments have shown that spatial representations of environments are enriched in place fields close to spots of interests (such as water pots \cite{Hollup01} or objects \cite{Bourboulou2019}) with respect to void regions. Numerical simulations reported in SM, Sec.~I.D\&E show that increasing the density of prescribed positions in regions of the physical space allows us to carve specific attractors in the neural activity space, representing preferentially those regions. This result is compatible with recent studies establishing the link between PF distribution and behavioral place preference  \cite{Mamad17}. Interestingly, our quenched PF theory can be applied to any particular set of PF, not necessarily homogeneously distributed over space; knowledge of the PF characteristics, {\em e.g.} from experimental measurements, allows us to determine the multispace correlation matrix $\boldsymbol{\cal C}$ in (\ref{matrixC}) and to make specific predictions. A proof of principle  of this approach is shown in SM, Sec.~II.D, where we compare the couplings found with SVM and with our quenched PF theory on synthetic data. 

Thirdly, the biological implications of our work remain to be worked out. Several improvements should be first brought in terms of biological plausibility. In particular one should consider continuous rather than binary neurons, explicitly distinguish excitatory and inhibitory neurons and impose Dale's law on the associated synapses, and take into account the sparse nature of synapses \cite{Guzman} and of place-cell activity \cite{Alme14} observed in CA3. Border effects, known to be important for hippocampal maps \cite{Barry06}, should also be considered instead of the simple periodic boundary conditions assumed here. Finally, it would be very interesting to study the dynamics of learning. As a preliminary attempt, we consider in SM, Sec.~I.F, an on-line version of the SVM learning algorithm \cite{adatron}, and show that the number of presentations of the patterns needed to stabilize a map is approximately proportional to $p$. Studying plausible learning rules could ultimately elucidate how the network progressively maturates to account for more and more fixed points and eventually defines a quasi-continuous attractor, as seems to be the case during the first weeks of development in rodents \cite{Dragoi}.

\vskip .2cm\noindent
{\bf Acknowledgements.} We are grateful to A. Treves for useful discussions. This work was funded by the HFSP  RGP0057/2016 project.

\end{document}

\begin{center}
  \LARGE Supplemental Material
\end{center}
\vskip .5cm
\title{Capacity-resolution trade-off in the optimal learning of multiple low-dimensional manifolds by attractor neural networks}
\vskip .3cm
\author{Aldo Battista}
\author{R\'emi Monasson}
\maketitle

\vskip 2cm

\tableofcontents

\newpage

\section{Numerical results: Support Vector Machine (SVM) learning}

\subsection{Dataset preparation: environments and distributions of PF}
We define an environment as a $D$-dimensional torus with unitary volume in which each neuron $i=1...N$ of the Continuous Attractor Neural Network (CANN) has a randomly located place field, {\em i.e.} a $D$-dimensional hyper-sphere of volume $\phi_0<1$, centered at position ${\bf r}_i$. We want to store in the network $L=\alpha N$ maps that differ through random rearrangements of the place-field (PF) center positions, ${\bf r}_i^\ell$, $\ell=1...L$ . Each map $\ell$ is approximated through a collection of $p$ random positions ${\bf r}^{\ell,\mu}$ in the environment; the maps become continuous in the large $p$ limit. For every position ${\bf r}^{\ell,\mu}$ we extract a pattern of activity of the network in the following way: all the neurons $i$ whose place field overlap with the position are active ($\sigma_i=1$), the others are silent ($\sigma_i=0$). A sketch representation of how we construct the patterns to store is drawn in the Fig.~1(b) of the main text. We end up with a data-set of $p\times L$ binary patterns $\{ \sigma_{i}^{\ell ,\mu} \}$, where $i$ is the neuron index, $\ell$ the environment index and $\mu$ the index of the position in map $\ell$.

\subsection{SVM learning procedure implementation}

After we have generated a data-set of activity patterns, we want to learn the connections $W_{ij}$ of the CANN that maximize the stability $\kappa$ at fixed $\alpha$ and $p$. This choice of the weights will ensure the biggest basins of attraction in the pattern space, {\em i.e.} robustness against thermal noise. In order to do that we will implement SVM learning \cite{SVM}. 
In practice, for each neuron $i$, we want to compute the connections $W_{ij}$ from the other neurons $j$ (with $W_{ii}=0$, no self-connection), which are solution of the following primal constrained convex optimization problem
\begin{equation}\label{primal}
\begin{aligned}
& \underset{\{W_{ij}\}}{\text{minimize}}
& & \frac{1}{2} \sum_{j (\ne i)} W_{ij}^{2} \, ,   \\
& \text{subject to}
& & (2\sigma_{i}^{\ell \mu}-1)\sum_{j (\ne i)} W_{ij}\, \sigma_{j}^{\ell \mu} \geq 1 , & \forall \, \ell, \mu  \, .
\end{aligned}
\end{equation}
We have to solve $N$ such problems to extract all the rows of the coupling matrix. The dual form of this problem is
\begin{equation}\label{qop}
\begin{aligned}
& \underset{\{\lambda_{\ell,\mu}\}}{\text{maximize}}
& & \sum_{\ell=1}^{L}\sum_{\mu=1}^{p} \lambda_{\ell \mu}-\frac{1}{2}\sum_{\ell,m=1}^{L}\sum_{\mu,\nu=1}^{p} (2\sigma_{i}^{\ell \mu} -1)(2\sigma_{i}^{m \nu} -1)\lambda_{\ell \mu} \lambda_{m \nu} \sum_{j(\neq i)}^{N} \sigma_{j}^{\ell \mu} \sigma_{j}^{m \nu} \, ,   \\
& \text{subject to}
& & \lambda_{\ell,\mu} \ge 0  , \, \forall \, \ell, \mu  \, , 
\end{aligned}
\end{equation}
where the $\lambda_{\ell \mu}$'s are Lagrange multipliers enforcing the constraints in (\ref{primal}). This optimization problem can be solved using available numerical routines \cite{convex}.
Once we obtain the $\lambda_{\ell \mu}$'s we can compute the connections through
\begin{equation}
W_{ij}=\sum_{\ell=1}^{L}\sum_{\mu=1}^{p}\lambda_{\ell \mu}\, (2 \sigma_{i}^{\ell \mu} -1)\, \sigma_{j}^{\ell \mu}  \, .
\end{equation}
We then normalize the rows of the couplings matrix to unity, {\em i.e.} $\sum_{j (\ne i)} W_{ij}^2=1$.
Finally, the stability $\kappa$ is computed through formula (3) of the main text. We have checked that the same values for $\kappa$  are obtained with a standard package for SVM, LinearSVC \cite{sklearn}. 

\begin{figure}[ht]
	\centering
	\includegraphics[width=.45\textwidth]{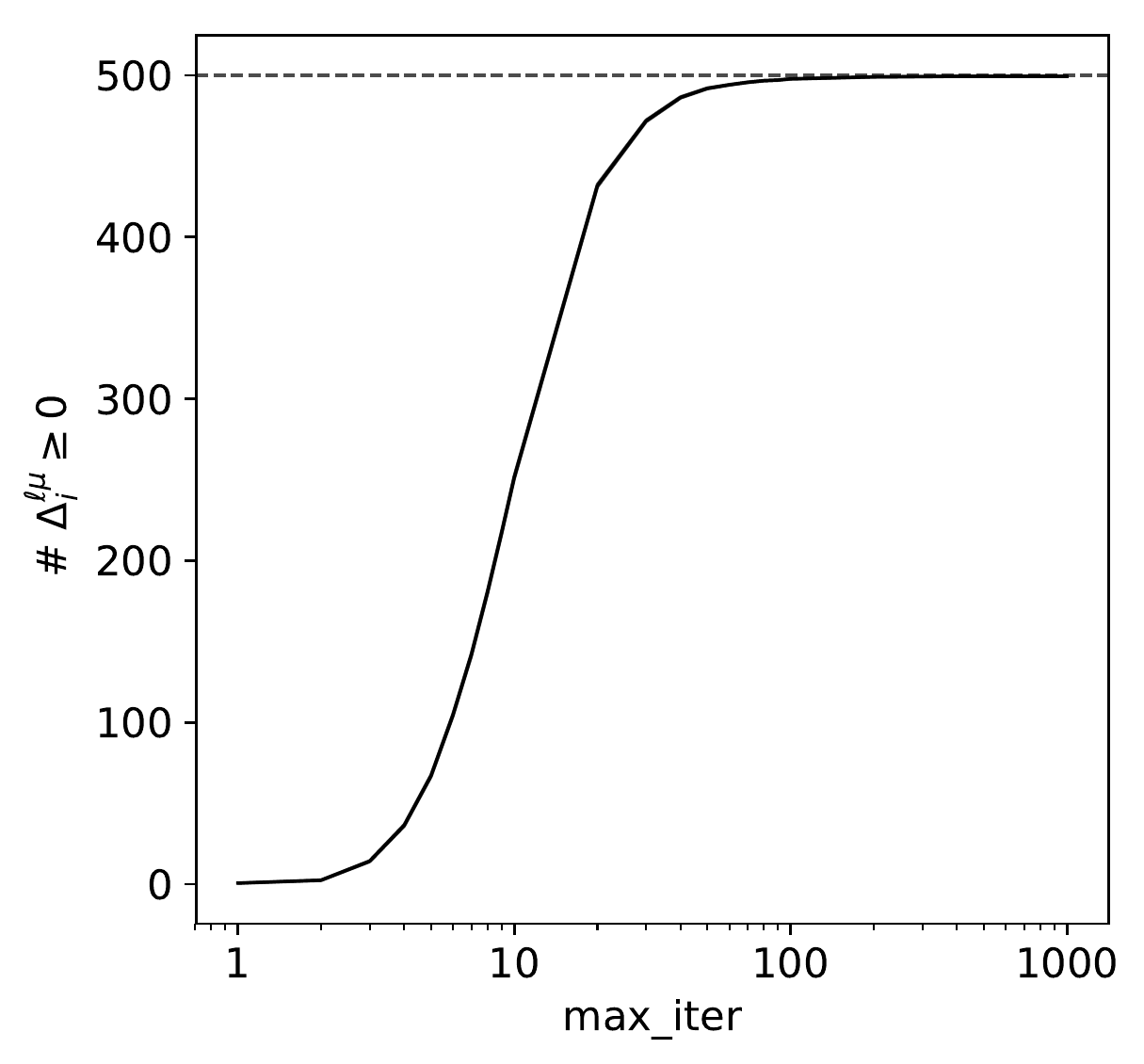}
	\caption{Number of patterns with positive stabilities ($y$-axis) vs. number of iterations of the quadratic optimization solver ($x$-axis) for one map ($L=1$) with $p=500$ points stored by a network with $N=1000$ neurons. Parameter values: $D=2$, $\phi_0=.3$.}
        \label{fig:FIG_IA}
\end{figure}

As an illustration of the learning procedure, we show in Fig.~\ref{fig:FIG_IA} how the number of stored patterns (with positive stabilities) grows as a function of the number of iterations of the quadratic optimization algorithm solving (\ref{qop}), until all $p$ prescribed patterns are stabilized.

\subsection{Couplings obtained by SVM}

\begin{figure}[ht]
	\centering
	\includegraphics[width=.8\textwidth]{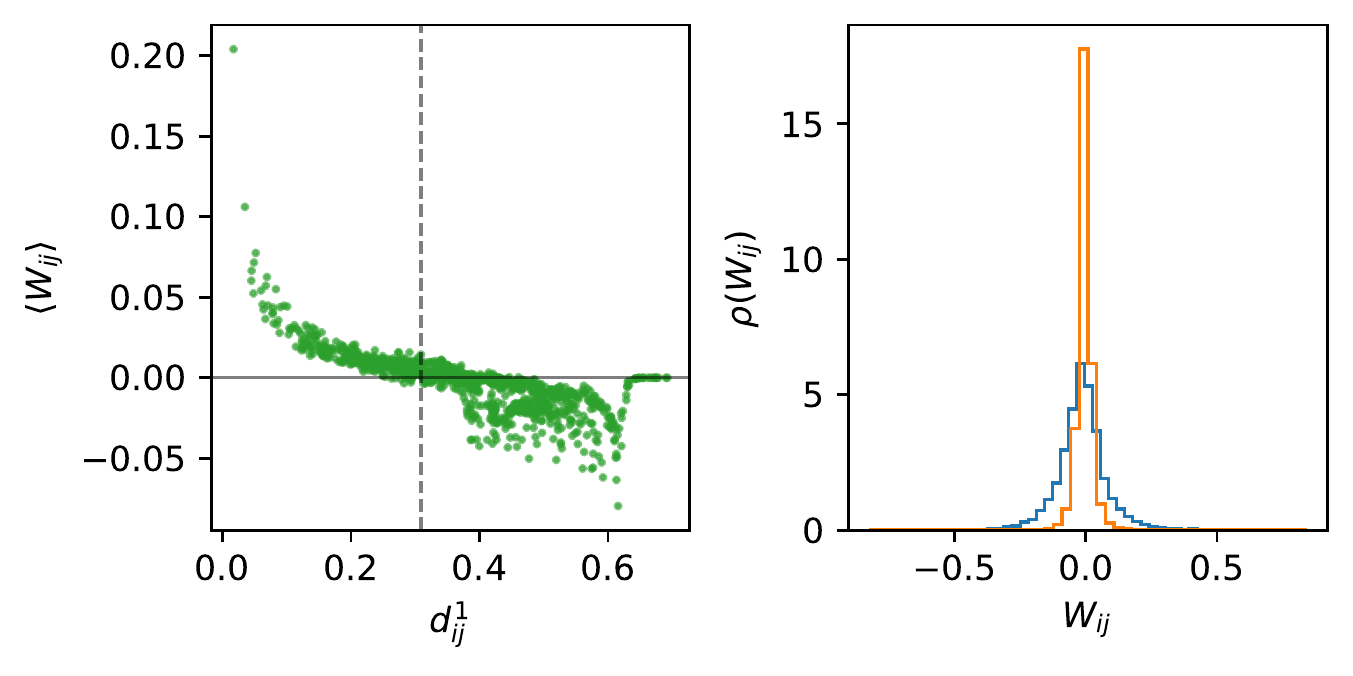}
	\includegraphics[width=.8\textwidth]{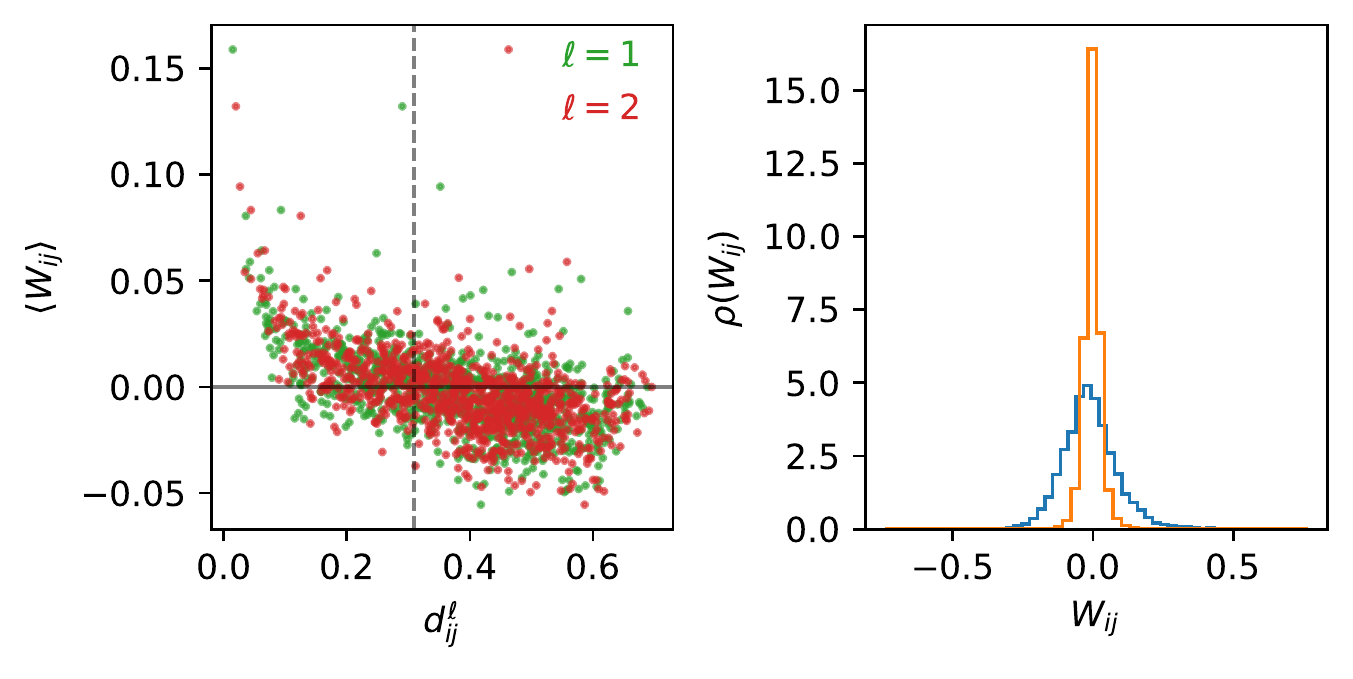}
	\caption{Couplings obtained after training with SVM for $L=1$ (top) and $L=2$ (bottom) maps. Left: dependence of the average coupling with the distance between the corresponding neurons; the vertical line locates the radius $r_c$ of the place fields. Averages were computed over $500$ samples of the $p$ positions per map at fixed PF centers; $N=1000$ neurons. Right: histograms of the couplings, for sizes $N=100$ (blue) and $N=1000$ (orange). Parameters: $D=2$, $ \phi_0=.3$ and $p\times L=N$. }
        \label{fig:FIG_IC}
\end{figure}

Hereafter, we report some qualitative features of the couplings obtained by SVM. As shown in Fig.~\ref{fig:FIG_IC} the couplings $W_{ij}$ are correlated with the distances $d_{ij}^{\ell} = |{\bf r}_i^\ell - {\bf r}_j^\ell|$ between the PF centers of the neurons $i$ and $j$ in the different maps $\ell$. Note that the dependence on distance is less marked as the number $L$ of maps increases, due to the interferences between the maps. 

In order to sustain a bump state with average activity $\phi_0$, couplings are excitatory at short distances, up to roughly the radius $r_c$ of the PF, and inhibitory at longer ones. The sign of the couplings can be intuitively understood. Two neurons at short distances have largely overlapping PF: their activities are likely to be equal, and having a large coupling helps increasing the stability, see eqn (3) in main text. If the distance is bigger than $r_c$,  the activities are likely to be different, hence inhibitory (negative) couplings will increase the stability. %All of this is due to the fact that SVM find the optimal couplings that are able to classify the state of a reference PFC given the status of the others.

Histograms of couplings in Fig.~\ref{fig:FIG_IC} (right) show that the amplitudes decay with $N$. In agreement with \cite{monasson} we expect the average values and standard deviations to scale, respectively, as $1/N$ and $1/\sqrt N$. 
 
\subsection{Miscellaneous results on optimal stability and capacity}

\subsubsection{Comparison with Hebb rule}

\begin{figure}[t]
	\centering
	\includegraphics[width=.4\textwidth]{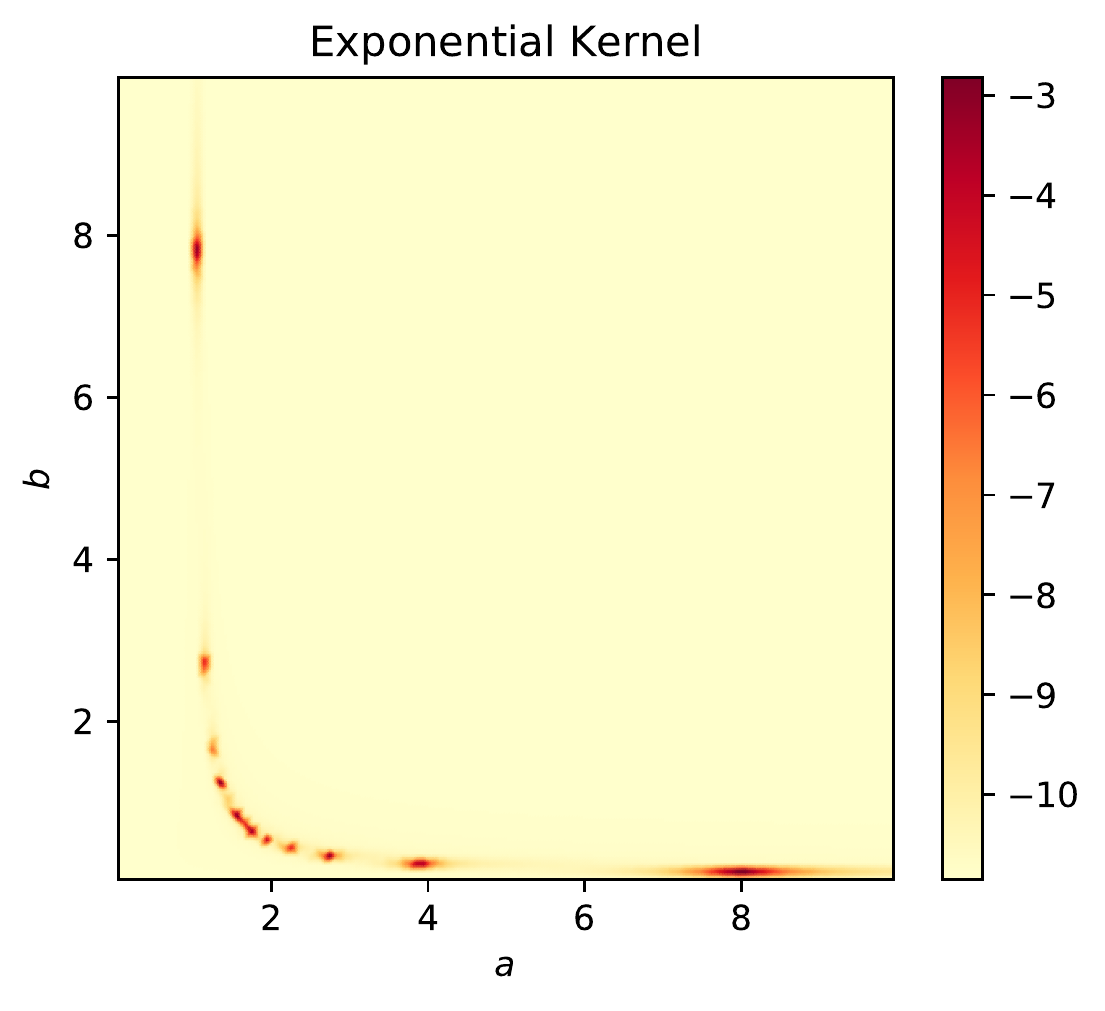} \hskip 1cm
	\includegraphics[width=.4\textwidth]{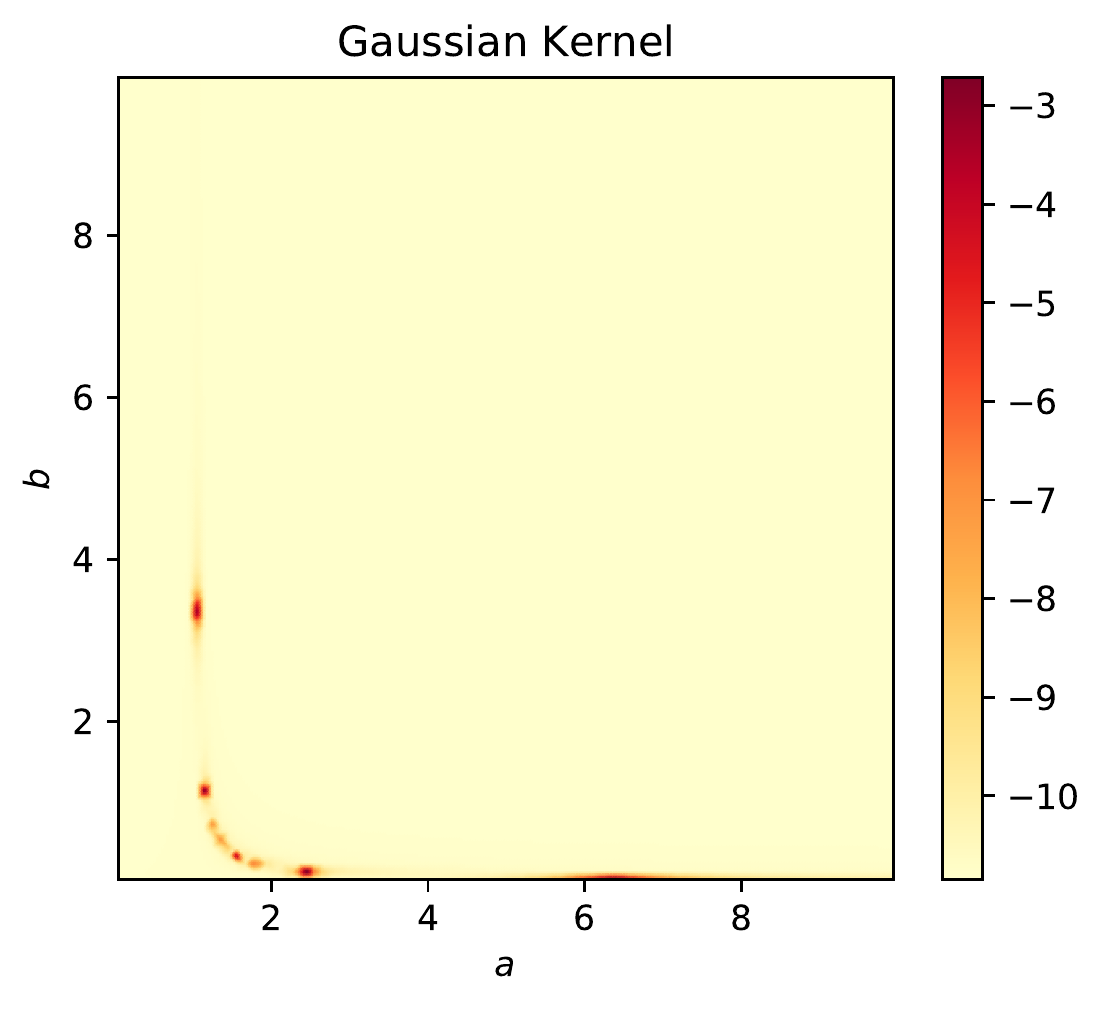}
	\caption{Stabilities obtained with the Hebb rule with exponential (left) and Gaussian (right) kernels on a given representative sample. The kernel parameters $a$ and $b$ vary from $0$ to $10$ with a step of $.01$. Parameter values: $N=1000$, $D=2$, $\phi_0=.3$, $\alpha=.1$ and $p=5$. The optimal value of the stability on that sample obtained by SVM is $\kappa \simeq .55$.}
        \label{fig:FIG_IE}
\end{figure}

Here we show that, as it should be by construction,
the stability obtained by SVM is always much higher than the one obtained by the Hebb rule $(2)$ defined in the main text. In order to do that we consider the cases of an exponential kernel,
\begin{equation}
w(d)=a \, e^{-d/b}-1 \ ,
\end{equation}
and of a Gaussian kernel, \begin{equation}
   w(d)=a \, e^{-d^2/b}-1 \ .
\end{equation}
We then optimize over $a$ and $b$; the value of the negative offset at large distance is arbitrary, since couplings are normalized row by row. Results for a typical sample are shown in Fig.~\ref{fig:FIG_IE}. The stability for the best kernel $w$ is always much lower (and negative in the examples considered here) than the optimal stability $\kappa$ found with SVM.

\subsubsection{Heterogeneous distribution of positions}

Throughout this work we have considered for simplicity that the $p$ positions were drawn uniformly at random to produce statistically homogeneous maps, {\em i.e.} without preferred positions. It is straightforward to extend this setting to the case of heterogeneous densities of prescribed positions.

\begin{figure}[ht]
	\centering
	\includegraphics[width=.497\textwidth]{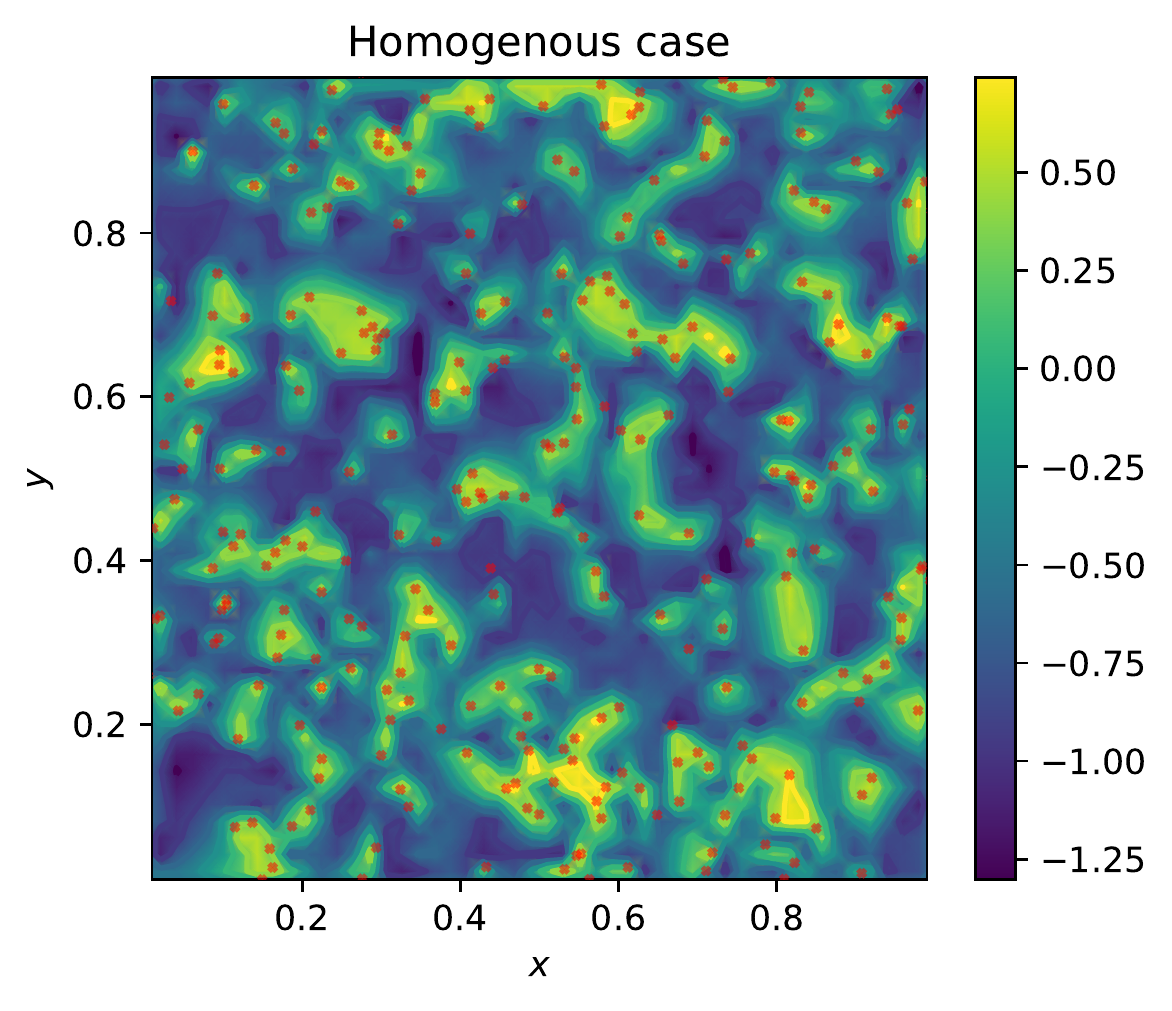}
	\includegraphics[width=.497\textwidth]{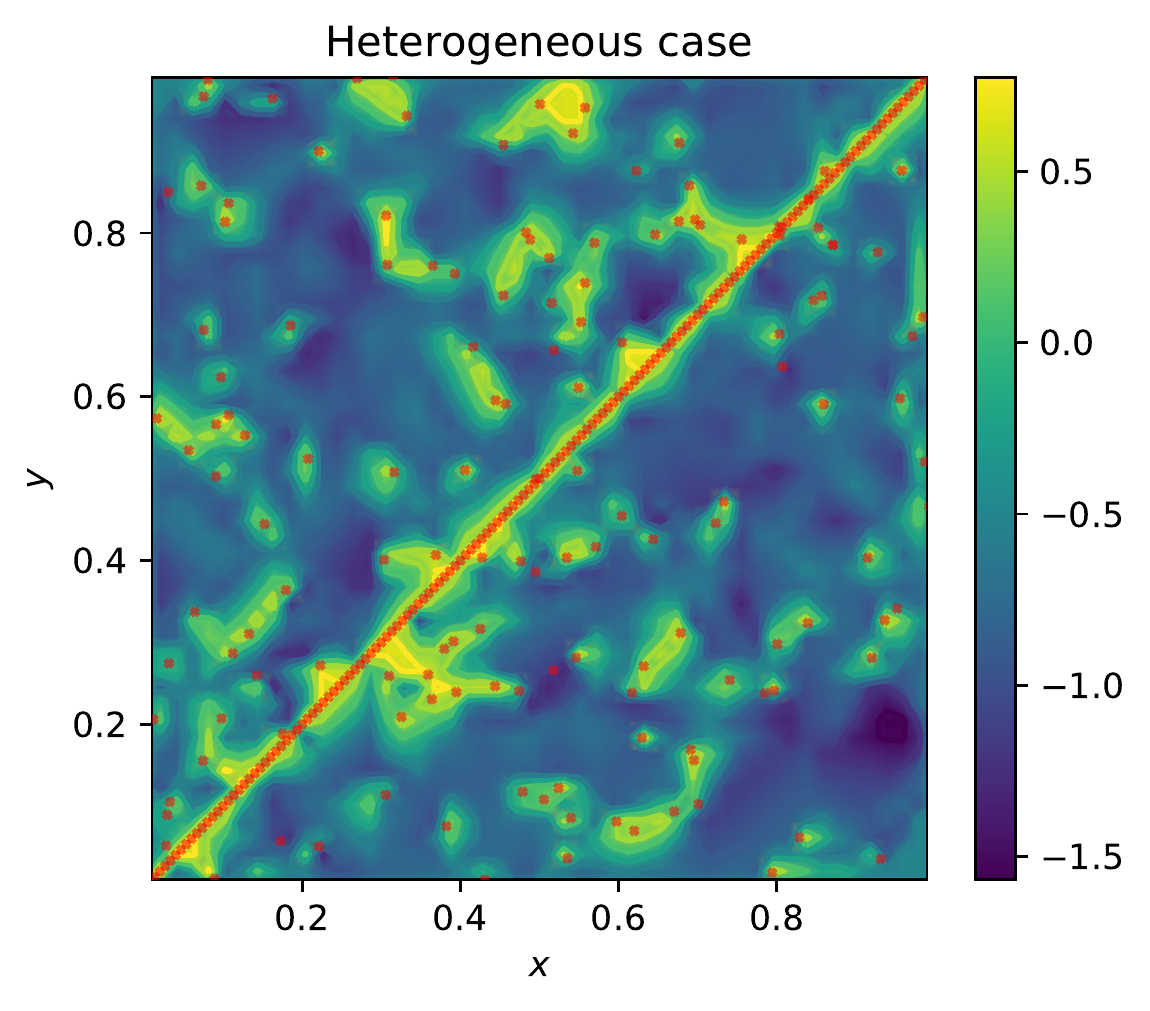}
	\caption{Distribution of local stabilities after the learning of a map with SVMs. Left) Homogeneous case: the $p$ positions of the data-set are drawn randomly. Right) Heterogeneous case: here the data-set has $150$ positions on the diagonal of the maps and the other $150$ positions are drawn at random. Here, $D=2$, $\phi_0=.3$, $N=1000$, $p=300$ and $L=1$. We show the contour map made from $2500$ realization of random positions, for which we evaluate the stabilities of the corresponding patterns. The overall network stabilities (minimal pattern stabilities) in the  homogeneous and heterogeneous cases are, respectively, $\kappa\simeq .44$ and $\kappa\simeq .53$ for the samples considered here.}
        \label{fig:FIG_IF}
\end{figure}

Figure~\ref{fig:FIG_IF} show the spatial distribution of stabilities for homogeneously scattered points (left) and a heterogeneous repartition on points, densely packed in a subregion (diagonal). In the latter case,  the strong heterogeneity in the local distribution of stabilities will favor the location of the bump along the zones with a major density of positions. As a consequence, a  $1D$-attractor is effectively built in the $D=2$-dimensional space, see videos described in Section I.E.3. 

\subsubsection{Dependence on $\phi_0$ and $D$}

\begin{figure}[h]
	\centering
	\includegraphics[width=.85\textwidth]{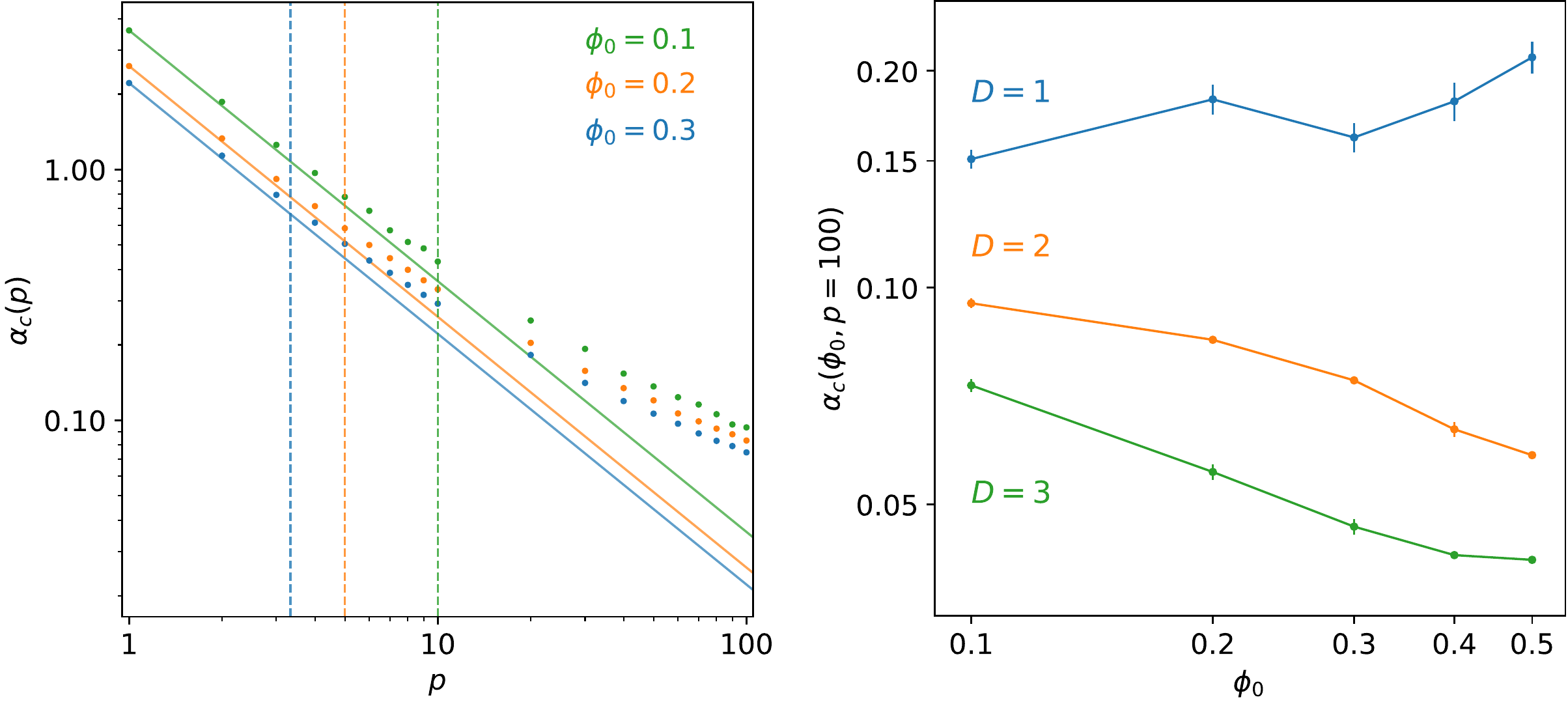}
	\caption{(Left) Scaling cross-over of $\alpha_{c}(p)$ vs. $p$ for different values of $\phi_0$. The vertical lines correspond to the values of $p_{c.o.} \sim \frac{1}{\phi_0}$.  We use for this results $D=2$, $N=5000$, and we have averaged over $50$ different realization of the environments and different realizations of the $p$ positions. (Right) Critical capacity obtained by SVM vs. $\phi_0$ for different values of $D$ in log-log scale. Parameter values: $N=5000$, $p=100$, Samples$=25$.}
        \label{fig:FIG_IG}
\end{figure}

Figure~\ref{fig:FIG_IG}(left) indicates that the behaviour of $\alpha_{c}(p)$ vs. $p$ changes from a $\frac{1}{p}$-scaling to a slower decay at a cross-over value  $p_{c.o.}\simeq \frac{1}{\phi_0}$. 

Figure~\ref{fig:FIG_IG}(right) shows that the behaviour of $\alpha_c(p)$ with $\phi_0$, obtained from SVM, is in qualitative agreement with equation (10) in the main text. In particular, we see that the critical capacity is largely independent of $\phi_0$ in dimension $D=1$, while it increases as the PF size $\phi_0$ shrinks in dimensions $D=2$, and even more so for $D=3$. Notice that the agreement with the asymptotic result given in equation (10) of the main text is not perfect here, due to the moderate value of the number of points in simulations ($p=100$). 

\begin{figure}[t]
	\centering
	\includegraphics[width=.45\textwidth]{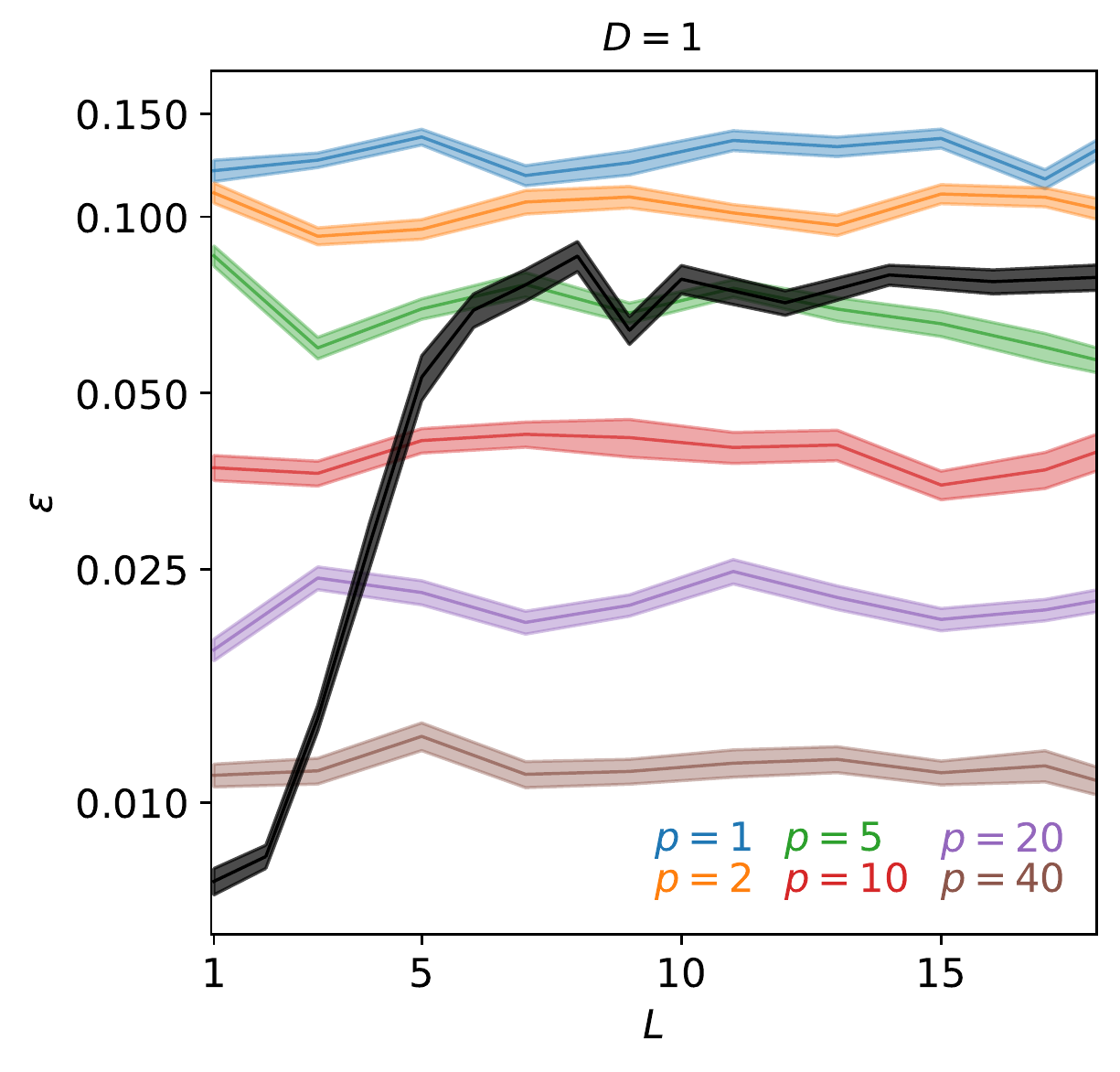}
	\includegraphics[width=.45\textwidth]{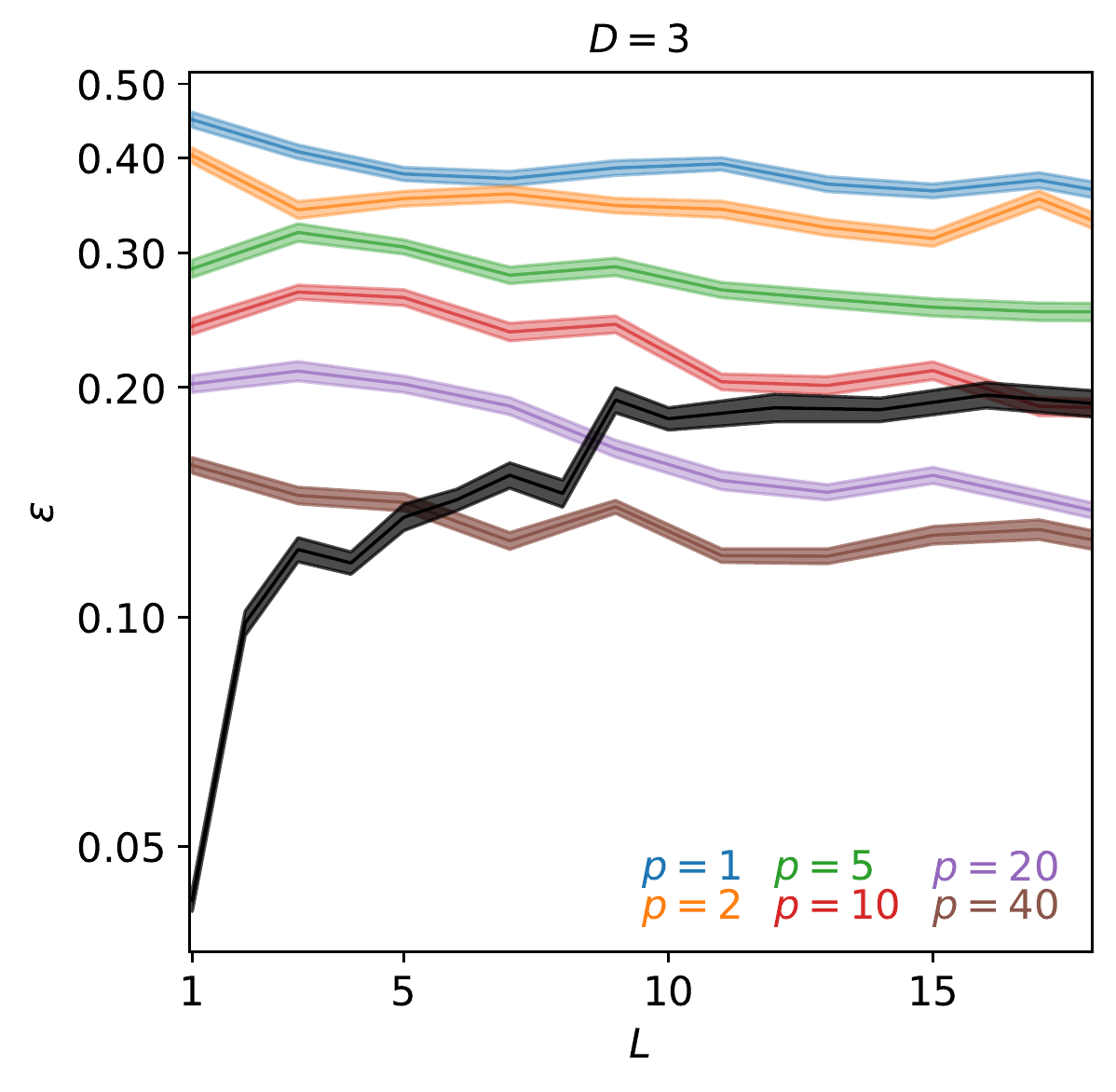}
	\includegraphics[width=.45\textwidth]{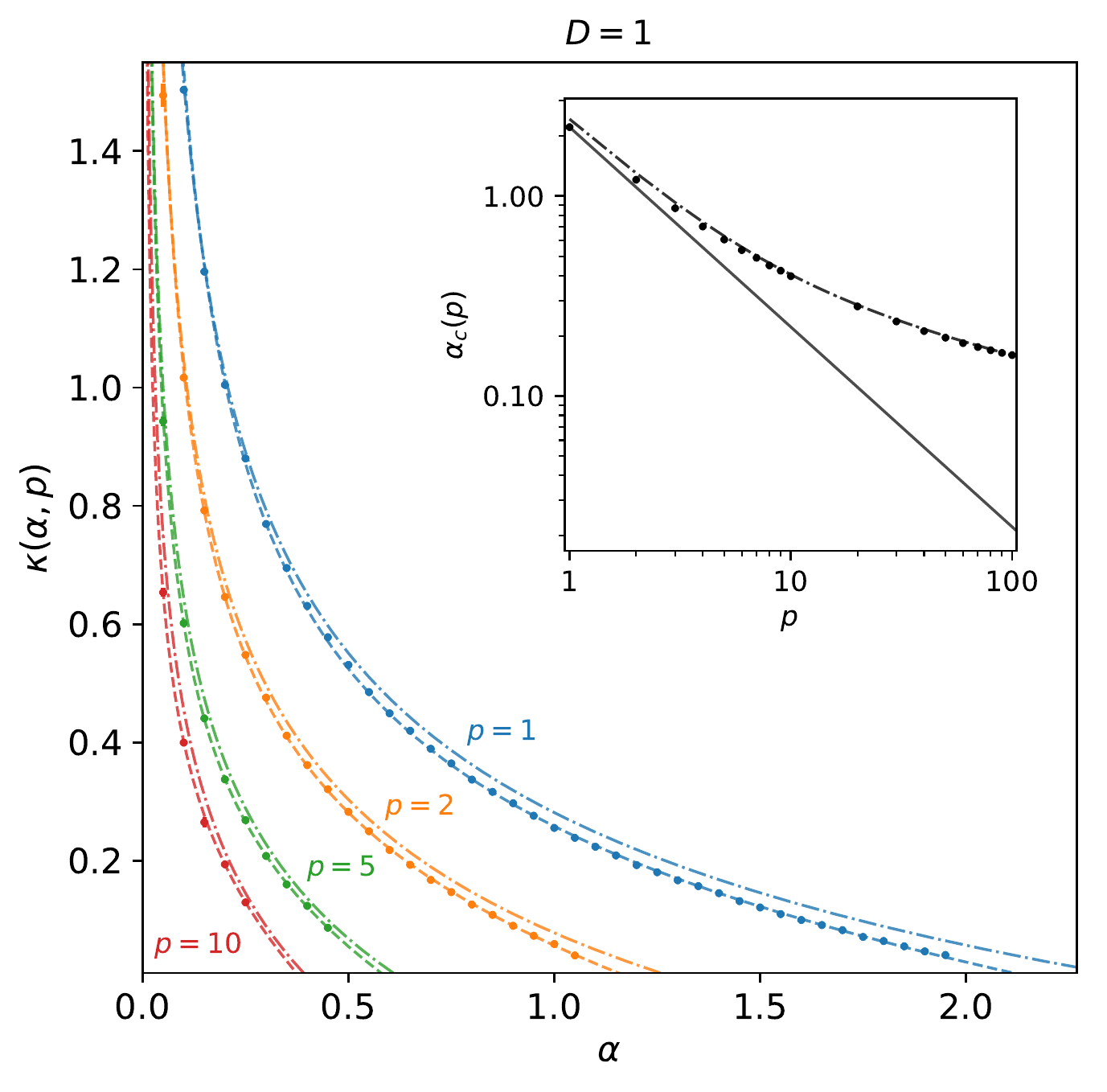}
	\includegraphics[width=.45\textwidth]{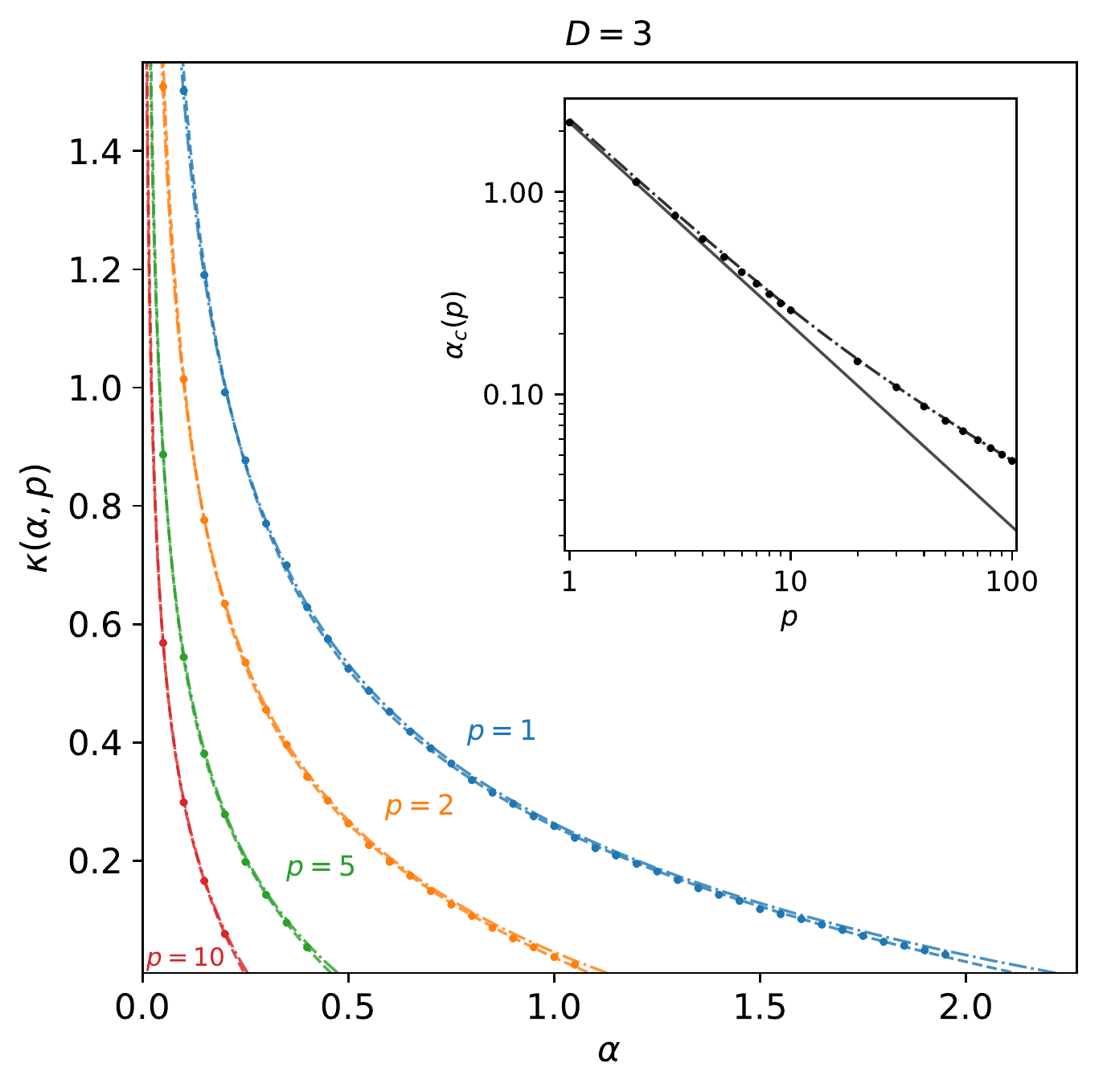}
	\caption{Same as Fig.~2(a) and Fig.~3 in the main text for dimensions $D=1$ (left) and $D=3$ (right).}
      \label{fig:FIG_IG2}
\end{figure}

Last of all, Fig.~\ref{fig:FIG_IG2} shows the spatial error of trained recurrent neural network and the optimal stability $\kappa$ vs. the load $\alpha$ for patterns generated in dimensions $D=1$ and $3$, completing the results shown for $D=2$ in the main text. We observe the faster decay of the critical capacity predicted by equation (10) in the main text with increasing values of $D$.

In these figures, as well as in the ones shown in the main text, the critical capacity $\alpha_c(p)$ was estimated as follows from SVM data. We estimated the optimal stabilities $\kappa$ (at fixed $p$) for $M$ different values of the load $\alpha$, with $M$ generally equal to 20. Then we fitted these points with the empirical function (depending on the parameters $a,b,c$)
\begin{equation}\label{kappetit}
    \kappa=\frac{a}{\sqrt{\alpha}}+b\, \alpha +c \ ,
\end{equation}
and extrapolated from the fit the value of the load at which the fitted function vanished;  this defined our estimate for $\alpha_c(p)$. Note that the small $\alpha$ behaviour in equation (\ref{kappetit}) above can be justified analytically from Gardner's calculation. 

\newpage

\subsection{Monte Carlo simulations}\label{mc}

Once the coupling matrix $W_{ij}$ has been learned, we may perform Monte Carlo simulations to investigate the behavior of the network.

\subsubsection{Zero temperature scheme ($T=0$)}
In order to compute the spatial error $\epsilon$ we consider that the dynamics of the system follows a sequential updating rule of the form
\begin{equation}\label{T0}
    \sigma_{i}(t+1)=\Theta \Big(\sum_{j(\neq i)}W_{ij}\, \sigma_{j}(t) \Big) \, ,
\end{equation}
where $\Theta$ is the Heaviside step-function and at every time $t+1$ we choose uniformly at random the neuron $i$ to update. Starting from an initial activity configuration, we track the system dynamics for at most $N^2$ MC steps ($N$ sweeps), and retain the visited configuration with the minimum number of violated constraints, {\em i.e.} with the highest number of non-negative stabilities
\begin{equation}
\Delta_{i}=(2\sigma_{i}-1)\sum_{j(\neq i)}W_{ij}\,\sigma_{j}\ge 0 \, \, \, .
\end{equation}
The choice of $N$ sweeps as a maximal simulation time is empirical:  we do not find that significantly better results are obtained by increasing this bound.
Actually, the dynamics often ends up in a fixed point with $\Delta_i>0$ for all neurons $i$ in much less sweeps.

We generate $L$ environments and $p$ points in each of them, learn the coupling matrix corresponding to these $p\times L$ patterns.  We then pick at random a position in one of the learned maps, and use that position to construct the initial activity configuration of the dynamics. After the dynamics described above is done we keep the final configuration and use it to decode the final position on that map, as the center of mass of PF (on the map) of the active neurons in the final configuration. The distance between this estimated position and the initial one (taking care of the periodic boundary conditions), after averaging over many initial positions ($100$ in the figures showed), defines the spatial error $\epsilon$.

%For how to concern the criteria for stopping the curves for SVMs, once we have fixed $p$, we continue to increase $L$ until the probability of learning all the $pL$ patterns is above a threshold, say around $70\%$.
\subsubsection{Finite temperature scheme ($T>0$)}

In order to show the diffusion of the activity bump within a map and the transitions between maps, we implement a noisy dynamical scheme, where neuronal states are updated stochastically according to the probabilities
\begin{equation}\label{Tn0}
   \Prob \Big( \sigma_{i}(t+1)  | \{ \sigma_j(t)\} \Big)=\frac{1}{1+\exp{\bigg[-\frac 1T \big( 2 \sigma_i(t+1)-1\big)\displaystyle{\sum_{j(\neq i)}W_{ij}\, \sigma_{j}(t)} }\bigg] } \, .
\end{equation}
The  bump of activity may form and sustain itself when  $T$ is comparable to, or smaller than the stability $\kappa$ of the network.

\subsubsection{Description of videos}

We illustrate the dynamical properties of the model with three examples:
\begin{itemize}
\item First, we consider a network with $N=1000$ neurons, in which we store one map ($L=1$) in dimension $D=2$ and with average activity $\phi_0=.3$. We consider than the case in which the learning is done on a small number of points, $p=30$, resulting in a large value of the stability, $\kappa=1.7$. And then the case in which $p=300$ is higher, and the stability is smaller: $\kappa=.6$. Our noise parameter $T$ is set to $.8$ to allow the bump to form in both cases. In the large $\kappa$ case,  the bump gets stuck very quickly in one of the $p$ training positions, depending on the initial configuration, see attached videos \texttt{LargeKappaL1.mp4} and \texttt{LargeKappaL1Bis.mp4}. In the small $\kappa$ case, the bump diffuses on the map, see attached video \texttt{SmallKappaL1.mp4}. For larger $p$, the bump can easily travel through the environment, with a large diffusion coefficient; in contrast, in the small $p$ case, the stability landscape is very rough and the bump is stucked close to the stored positions. 
\item In the second example we consider the case of $L=2$ maps and $p=150$ points. The other parameters have the same values as in the first example, {\em e.g.} the stability is fixed to $\kappa=.5$. In the video \texttt{SmallKappaL2.mp4} we see that, as $\kappa$ is small, the bump diffuses in one maps and sporadically jumps to the other map.
\item The third example corresponds to the heterogeneous distribution of positions shown in Fig.~\ref{fig:FIG_IF}, right. The  video \texttt{SmallKappaL1Hetero.mkv} was obtained with the same parameters as in \texttt{SmallKappaL1.mp4}, but with $150$ out of the $p=300$ positions drawn on the diagonal of the map; the stability of the network was $\kappa=.5$.  
\end{itemize}

\subsection{On-line version of the SVM learning algorithm}

\begin{figure}[ht]
	\centering
	\includegraphics[width=1\textwidth]{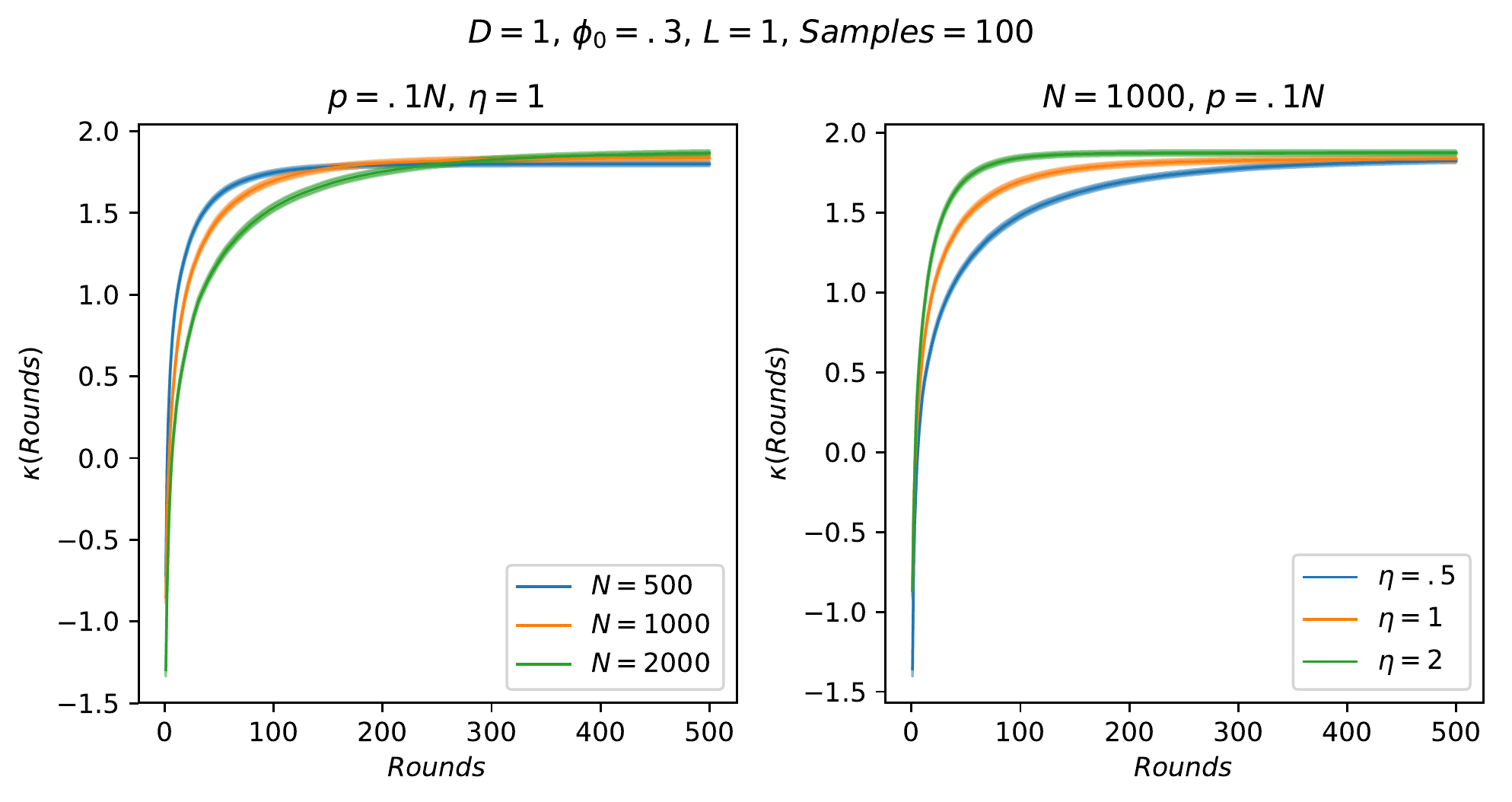}
	\caption{Stability of the network $\kappa$ vs. the number of rounds in the case of a single map for different values of $N$, $p$ and $\eta$, see figure. Results are averaged over 100 samples.}
	\label{fig:Learning_Dynamics}
\end{figure}

The SVM algorithm exposed in the above section is off-line: all the patterns are available to the learning procedure at all times, which is not biologically realistic.  As a preliminary attempt to understand how maps are learned, we have implemented an on-line learning algorithm, which is an adaptive version of the perceptron algorithm, see \cite{adatron}. In this procedure,  patterns are presented one after the other. We may choose the order of presentation, as well as the learning rate $\eta$. In order to study the time needed for the algorithm to store a map, we have run the on-line learning algorithm in the simplest case of a single environment ($L=1$). We have monitored the stability of the network during this learning phase as a function of the number of training rounds, a round corresponding here to the  presentation of all the patterns in the dataset. In Fig.~\ref{fig:Learning_Dynamics}, we show that the number of rounds needed to stabilize a map is roughly proportional to $p/\eta$ (for a fixed ratio $p/N$). 

It would be interesting to relate this finding to biological results. Let us remark that the presentation of repeated rounds considered here could be realistic for an animal exploring the same environment several times, in particular a 1D corridor in which the sequence of visual inputs remains roughly unchanged from one exploration to another. Experiments show that changes in the environment (insertion of one object) lead to the production of a new representation, which is stabilized over 4-5 explorations, see Fig.~4J in \cite{bourboulou}. 

\section{Theoretical results: Statistical physics of optimal CANN}

\subsection{Gardner's framework for CANN} \label{gard}

Here we are going to extend the Gardner theory for capacity of the perceptron (SVM with linear kernel and hard margin) \cite{gardner} to the case of continuous attractors. The training set consist of $p \times L$ binary patterns $\{ \sigma_{i}^{\ell \mu} \}$ constructed by drawing randomly $p$ positions in each of the $L$ environments so that the resulting patterns are spatially correlated. The stability of the $i$ component of the pattern that correspond to position $\mu$ in the environment $\ell$ is given by
\begin{equation}
\Delta_{i}^{\ell \mu}=(2\sigma_{i}^{\ell \mu}-1)\sum_{j(\neq i)}W_{ij}\,\sigma_{j}^{\ell \mu} \, .
\end{equation}
The training set is said to be stored if all the patterns have stabilities larger than some threshold $\kappa \geq 0$. 

The volume in the space of couplings that corresponds to admissible solutions of the storage problem, is
\begin{equation}
     Z=\int \prod_{i\ne j}^{N} dW_{ij} \prod_{i} \delta\bigg(\sum_{j(\neq i)}W_{ij}^2-1\bigg) \prod_{i, \ell, \mu} \theta \Big((2\sigma_i^{\ell \mu} -1)\sum_{j (\ne i)} W_{ij} \,\sigma_{j}^{\ell \mu}-\kappa \Big)
\end{equation}
and is equal to the product of the $N$ single-site volumes $Z_{i}$, with $i=1...N$. So we may focus for example on the volume associated with $i=1$:
\begin{equation}
    Z_{1}=\int \prod_{j=2}^{N} dW_{j}\; \delta \bigg(\sum_{j\ge 2} W_j^2-1 \bigg) \prod_{\ell, \mu} \theta \Big((2\sigma_1^{\ell \mu} -1)\sum_{j\ge 2} W_j \,\sigma_{j}^{\ell \mu}-\kappa \Big),
\end{equation}
where $W_j\equiv W_{1j}$. Using the replica method \cite{spinglass}, we compute the average of $\log Z_1$ over the patterns. Introducing integral representations of the Heaviside function and exploiting the statistical independence of the different maps, we write the average of the $n^{th}$ power of the volume,
\begin{equation}\label{z1n}
    \langle Z_{1}^{n} \rangle = \int \prod_{j,a} dW_{ja}  \prod_{a} \delta \Big(\sum_{j} W_{ja}^2-1 \Big)\; \chi({\bf W})^{\alpha N}
\end{equation}
with $a=1,...,n$ is the replica index, and
\begin{equation}\label{chiw}
    \chi ({\bf W})=\int \prod_{\mu=1}^p d {\bf \hat  r}_{\mu} \int \prod_{j=1}^N d {\bf r}_{j}\int_\kappa^\infty \prod_{\mu, a} dt_{\mu a}  \int _{-\infty}^\infty \prod_{\mu, a} \frac{d \hat t_{\mu a}}{2 \pi} e^{i \sum_{\mu, a}\hat t_{\mu a} t_{\mu a}} \prod_{j}e^{-i\sum_{\mu, a} \hat t_{\mu a}(2\sigma_1^{\mu}-1)W_{ja}\sigma_j^\mu} \, ,
\end{equation}
where ${\bf \hat r}_\mu$ denotes the $p$ prescribed locations in the environment, and ${\bf r}_j$ the $N$ PF of the neurons in the map.  We first carry out explicitly the integrals over the PF with indices $j=2,3,...,N$, leaving the integrals over ${\bf r}_1$ and all ${\bf \hat r}_\mu$ in $\chi ({\bf W})$. We introduce the order parameters
\begin{equation}
    m^{a}=\phi_0 \, \sum_{{j\ge 2}}W_{ja}
\end{equation}
and 
\begin{equation}
    q^{a b} =\sum_{{j\ge 2}} W_{j a} W_{j b} \ ,
\end{equation}
and rewrite
\begin{equation}
\begin{split}
    \langle Z_{1}^{n} \rangle = \int \prod_{j,a} dW_{j,a} \int \prod_{a} \frac{d \hat u^{a}}{4 \pi} e^{\sum_{a}\frac{\hat u^{a}}{2}(1-\sum_j W_{ja}^2)}  \int \prod_{a} \frac{d\hat m^{a} d m^{a}}{2 \pi}e^{\sum_{a}\hat m^{a}(m^{a}-\phi_0\sum_j W_{ja})} \\
\times    \int \prod_{a \le b}\frac{d \hat q^{ab} d q^{ab}}{2 \pi}e^{\sum_{a \le b}\hat q^{ab}(q^{ab}-\sum_j W_{ja}W_{jb})} \, \chi({\bf W})^{\alpha N} 
\end{split} 
\end{equation}
where we have used the integral representation of the Dirac-delta function.
Let $\Phi({\bf r})$ be the indicator function of the place field centered in $\bf 0$: $\Phi=1$ if $|{\bf  r}|<r_c$, where $r_c$ is the radius of the PF (with $\int d{\bf r} \, \Phi({\bf r})=\phi_0$), and 0 otherwise. Let $\Gamma({\bf r}) = \int d{\bf r}' \Phi({\bf  r}')\,\Phi({\bf r}-{\bf r}')$ the correlation function of $\Phi$.
Given $p$ points ${\bf \hat  r}_\mu$, $\mu=1,...p$ drawn uniformly at random in space, we define the $p\times p$  Euclidean random matrix with entries
\begin{equation}
\boldsymbol\Gamma_{\mu, \nu} \big(\hat {\cal R}\equiv\{{\bf \hat  r}_\mu\}\big) = \Gamma\big( {\bf \hat r}_\mu - {\bf \hat r}_{\nu}\big) - \phi_0^2 \ .
\end{equation}
We can rewrite $\chi$ as
\begin{equation}
\begin{split}
    \chi({\bf W})=\int \prod_{\mu} d {\bf \hat r}_{\mu}  \int d {\bf r}_{1} \int_{\kappa}^{\infty}\prod_{\mu, a} \frac{dt_{\mu a}}{\sqrt{2\pi}} \int_{-\infty}^{\infty}\prod_{\mu, a} \frac{d\hat t_{\mu a}}{\sqrt{2\pi}}e^{-\frac{1}{2}\sum_{\mu,\nu,a,b}q^{ab}\boldsymbol\Gamma_{\mu,\nu} (\hat {\cal R})\hat t_{\mu a} \hat t_{\nu b}} \\
    e^{-i \sum_{\mu, a} m^{a} \hat t_{\mu a} \Phi({\bf r}_{1}- {\bf \hat  r}_{\mu})+i\sum_{\mu,a}\hat t_{\mu a} t_{\mu a}} \, .
\end{split}
\end{equation}
Due to translation invariance,  the integral over ${\bf r}_{1}$ is irrelevant,  and we can set ${\bf r}_{1}= {\bf 0}$. We can now make the RS Ansatz  (expected to be valid since the domain of suitable couplings is convex) on the structure of the order parameters and their conjugate variables, and, after standard manipulation, we write the $n^{th}$ power of the volume  in the small $n$ limit as
\begin{eqnarray} \label{bracket}
    \frac{\langle Z_{1}^{n} \rangle-1}{n N} &\simeq& \frac{1}{2 \epsilon} \Bigg \{1-\alpha \int \prod_{\mu} d {\bf \hat r}_{\mu} \int \prod_{\mu} \frac{d z_{\mu}}{\sqrt{2 \pi}}\frac{ \exp\big(-\frac{1}{2}\sum_{\mu,\nu}z_{\mu}\boldsymbol\Gamma(\hat {\cal R})_{\mu,\nu}^{-1}z_{\nu} \big)}{\sqrt{\det \boldsymbol\Gamma (\hat {\cal R})}}  \\ 
    &\times & \min_{\{t_{\mu}\ge  \kappa +m\}} \sum_{\mu,\nu}\big[t_{\mu}-(z_{\mu}+2\, m\,\Phi({\bf r}_{1}- {\bf \hat  r}_{\mu})\big]\boldsymbol\Gamma(\hat {\cal R})_{\mu,\nu}^{-1}\big[t_{\nu}-(z_{\nu}+2\,m\,\Phi({\bf r}_{1}- {\bf \hat  r}_{\nu})\big] \Bigg \} + O \Bigg( \Big|\log  \frac 1\epsilon\Big|\Bigg) \ , \nonumber
\end{eqnarray}
where we have computed the integrals over the order parameters and related conjugate variables thanks to the saddle-point method. Since we are interested in the critical capacity we have also restricted our analysis to the case $\epsilon=1-q\ll 1$,  in which the space of solutions reduces to the optimal coupling matrix.
We finally obtain the expression for the critical capacity $\alpha_c(\kappa ,p) = \displaystyle{ \max _m \alpha_{c}(m;\kappa,p)}$, where $\alpha_{c}(m;\kappa,p)$ is the load $\alpha$ cancelling the terms inside the curly brackets in (\ref{bracket}).

\subsection{Case of a single location per map ($p=1$)}

We now show that, when a single pattern is present in each map ($p=1$), the equations above are equivalent to the celebrated Gardner calculation in the case of biased patterns \cite{gardner}. 

For $p=1$, the Euclidean random matrix ${\cal C}$ reduces to the scalar
\begin{equation}
{\cal C}_{1,1}=\phi_0(1 - \phi_0)\equiv\frac{1-M^{2}}{4} \ ,
\end{equation}
where $M$ is the average activity of the binary pattern in $\pm 1$ notations, {\em i.e.} under the change of variable $\Phi({\bf r}_1-{\bf \hat r}_1)=\{0,1\}\to\xi=\{-1,+1\}$. The convex optimization problem to be solved in (\ref{bracket}) thus amounts to compute
\begin{equation}\label{t111}
 F(z_1,v,\kappa) = \min _{\{t_1 \ge \kappa\}} \bigg[ \dfrac{4}{1-M^{2}} \big( t-(z_1+v\, \xi)\big)^{2}\bigg]\ ,
\end{equation}
where $v=m\,\phi_0$ and the Gaussian variable $z_1$ in (\ref{bracket}) has zero mean and variance ${\cal C}_{1,1}$.
The minimum over $t_1$ in (\ref{t111}) can easily be determined, with the result
\begin{equation}
F(z_1,v,\kappa) = 
\begin{cases}
     \dfrac{4}{1-M^{2}} \big( \kappa-(z_1+v\, \xi)\big)^{2}& \text{if } \kappa\geq z_1+v\, \xi\\
    0              & \text{otherwise}.
\end{cases}
\end{equation}
As ${\bf \hat r}_1$ is drawn uniformly at random, $\xi$ is a random binary variable: 
\begin{equation}
\xi = 
\begin{cases}
     +1& \text{with probability}\quad \dfrac{1+M}{2} \ , \\
    -1              & \text{with probability}\quad \dfrac{1-M}{2}.
\end{cases}
\end{equation}
We get, with the normalized Gaussian variable $z=z_1 \times 2/\sqrt{1-M^2}$ and the measure $Dz=dz/\sqrt{2\pi}\, \exp(-z^2/2)$,
\begin{equation}
\frac 1{\alpha_c (v;\kappa,p=1)} = \frac{1+M}{2}\int_{\frac{2v M-2\kappa}{\sqrt{1-M^{2}}}}^{\infty}Dz \bigg( \frac{2\kappa-2v M}{\sqrt{1-M^{2}}}+z\bigg)^{2}+ \frac{1-M}{2}\int_{\frac{-2\kappa-2v M}{\sqrt{1-M^{2}}}}^{\infty}Dz \bigg( \frac{2\kappa+ 2v M}{\sqrt{1-M^{2}}}+z\bigg)^{2}\ ,
\end{equation}
where $v$ is chosen in order to maximize $\alpha _c(v;\kappa,p=1)$:
\begin{equation}
\frac{1+M}{2}\int_{\frac{2v M-2\kappa}{\sqrt{1-M^{2}}}}^{\infty}Dz \bigg( \frac{2\kappa-2v M}{\sqrt{1-M^{2}}}+z\bigg)= \frac{1-M}{2}\int_{\frac{-2\kappa-2v M}{\sqrt{1-M^{2}}}}^{\infty}Dz \bigg( \frac{2\kappa+2 v M}{\sqrt{1-M^{2}}}+z\bigg).
\end{equation}
These equations coincide with the results of \cite{gardner} up to the change $\kappa\to 2 \kappa$ due to the fact that the neuron activities take here values 0,1 and not $\pm 1$.

\subsection{Gaussian theory with quenched PF}

In order to compute the critical capacity $\alpha_{c}(\kappa)$ with the approach of Section \ref{gard} we have to solve a $p$-dimensional constrained quadratic optimization problem, depending on $p$ correlated Gaussian random variables, see (\ref{bracket}), and then average over $p$ random positions. This task becomes quickly intractable in practice as  $p$ increases. In this section, following closely \cite{monasson}, we present an alternative approximate approach that allows us to reach arbitrarily large values of $p$. While this calculation is approximate, it is argued that it becomes exact in the large $p$ limit. A potentially interesting feature of this approach is that it holds at fixed PF, instead of averaging over them as in Section \ref{gard}, and could be applied to specific situations, {\em e.g.} sets of PF measured in experiments.

\subsubsection{Replica calculation} 

\begin{figure}[ht]
	\centering
	\includegraphics[width=1.\textwidth]{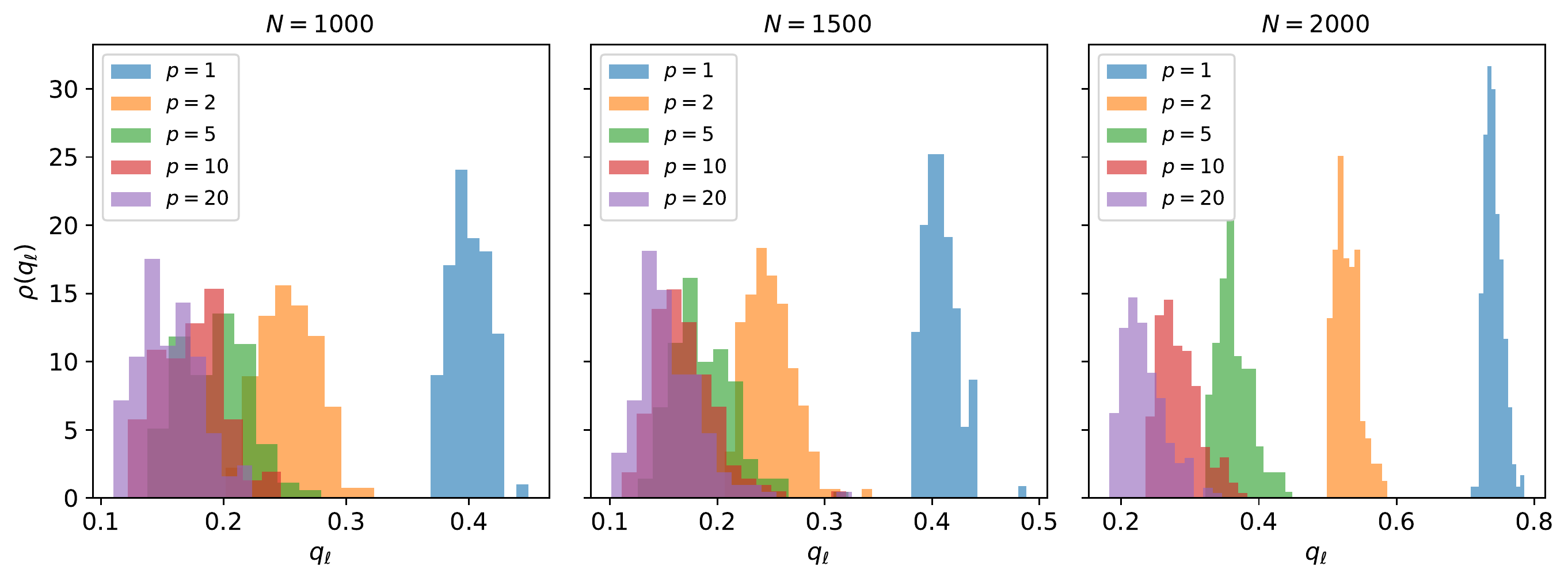}
	\caption{Distributions of the overlaps $q_{\ell}$ for different values of $N$ and $p$. It is clear that the histograms are roughly Gaussian. We use for this results $D=2$, $\phi_0=.3$, $\alpha=.1$,  and we have averaged over $500$ realization of the $p$ positions at fixed PF.}
        \label{fig:FIG_IIC}
\end{figure}

Starting from the replicated volume $\langle Z_1^{n}\rangle$ in (\ref{z1n}), we now perform first the average over each one of the $p$ locations in $\chi({\bf W})$ in (\ref{chiw}) as follows
\begin{equation}\label{expansion_hatt}
\int  d{\bf \hat r}_\mu\; \exp\left(-i\sum_{a} \hat t_{\mu a} (2\sigma_1^{\ell,\mu}-1)\sum _{j\ge 2} W_{ja}\,\sigma_j^{\ell,\mu} \right)= \exp\left(-i\sum_{a} m_{\ell}^{a}\, \hat t_{\mu a}-\frac{1}{2}\sum_{a,b}\hat t_{\mu a} (q_{\ell}^{ab}-m_{\ell}^{a}m_{\ell}^{b})\hat t_{\mu b} + O(\hat t ^3)\right)
\end{equation}
where we have reintroduced the map index $\ell$ to underline that the PF are kept fixed here. The order parameters in the formula above are
\begin{equation}
    m_{\ell}^{a}=\sum_{j\ge 2} W_{ja}\, \Big(2\, \Gamma\big(|{\bf r}_j^\ell-{\bf r}_1^\ell|\big)-\phi_{0}\Big)
\end{equation}
and
\begin{equation}
    q_{\ell}^{ab}=\sum_{j,k\ge 2}W_{j a}\, W_{kb}\;   \Gamma\big(|{\bf r}_j^\ell-{\bf r}_k^\ell |\big)  \ .
\end{equation}
\vskip .3cm \noindent
We simplify the calculation with two approximations:
\begin{itemize}
    \item We truncate the expansion in powers of $\hat t$ in (\ref{expansion_hatt}) to the second order, and omit all higher order terms. This amounts to approximate the distribution of couplings $W_{ij}$ (at fixed PF) by a Gaussian. This approximation is valid only if the couplings fluctuate weakly around their means, which is the case in the large-$p$ limit, see Section II.C.4.
    \item We also neglect the dependence of the order parameters $m_\ell$ and $q_\ell$ above on the map $\ell$. The histogram of the overlaps $q_\ell$ measured by SVM are shown in Fig.~\ref{fig:FIG_IIC}. As can be seen from the figure, the distribution of overlaps is not concentrated in the large-$N$ limit at fixed $p$. Therefore, while $m_\ell^{a}=m^a$ and $q_\ell^{ab}=q^{ab}$ is a valid Ansatz for the saddle-point equations of the log. partition function (due to the statistical equivalence between the maps), we expect Gaussian fluctuations to be relevant even in the infinite-$N$ limit.  However, as $p$ increases, these fluctuations are smaller and smaller, and are  asymptotically negligible. The order parameters then reduce to, after summation over the maps $\ell=1...L$,
 \begin{equation}
  m^a \equiv \frac 1L \sum_{\ell=1}^L  m_{\ell}^{a}=\sum_{j\ge 2} W_{ja}\, \Big(2\, \boldsymbol {\cal C} _{1j} \big(\{ {\bf r}_j^\ell  \}\big)-\phi_{0}\Big)
\end{equation}
and
\begin{equation}
q ^{ab} \equiv \frac 1L \sum_{\ell=1}^L     q_{\ell}^{ab}=\sum_{j,k\ge 2}W_{j a}\, W_{kb}\;  \boldsymbol {\cal C} _{jk}\big(\{ {\bf r}_j^\ell  \}\big) \ .
\end{equation}
The $N\times N$ multi-space Euclidean random matrix ${\boldsymbol{\cal C}}$ appearing in the expressions above is defined in equation (8) of the main text. In the following, we denote by $\rho (\lambda)$ the density of eigenvalues $\lambda$ of ${\boldsymbol {\cal C}}$. This density is self-averaging when the PF are randomly drawn in the large $L,N$ double limit. Its resolvent, defined as 
\begin{equation}\label{resolvent}
g(U) =\int d\lambda \, \frac{\rho(\lambda)}{\lambda+U} \ ,
\end{equation}
where the integral runs over the support of $\rho$, is solution of the implicit equation (9) of the main text \cite{battista19}.
\end{itemize}

Within the RS Ansatz, the overlap matrix $q^{ab}$ is fully characterized  by its diagonal and off-diagonal elements that we denote by, respectively, $s$ and $q$:
\begin{equation}\label{defsq}
s= \sum_{i,j\ge 2} \langle{\boldsymbol {\cal C}}_{ij} \, [ W_{1i}\,  W_{1j} ] \rangle\quad, \qquad q= \sum_{i,j\ge 2} \langle {\boldsymbol {\cal C}}_{ij} \,  [ W_{1i} ]\, [  W_{1j}] \rangle \ .
\end{equation}
where, as before, the brackets denotes the average over the random patterns, and the square parenthesis stand for the average over all couplings satisfying the inequalities (\ref{primal}).

Following closely \cite{monasson}, we obtain the expression of the average logarithm of the volume,
\begin{equation}\label{logvol}
\begin{split}
    \frac{\langle\log  Z_1 \rangle}{N}=-\frac{1}{2}q \hat q+ s \hat s + m \hat m + \hat u - \frac{1}{2}\int d\lambda \,\rho(\lambda)\bigg[ \log(2\hat u +(2\hat s - \hat q)\lambda)+\frac{\hat q \lambda}{2 \hat u+(2\hat s - \hat q) \lambda} \bigg] \\
    +\frac {\hat m^2 \, \Xi}{2 (2\hat s - \hat q)}+\alpha\,  p \,\int D z \log H\bigg(\frac{z\sqrt{q-m^2}-m+\kappa}{\sqrt{s-q}}\bigg)
\end{split}
\end{equation}
where $Dz$ denotes the Gaussian measure, $H(x)=\int _x ^\infty Dz=\frac{1}{2} \erfc (\frac{x}{\sqrt{2}})$, and the $\hat \cdot$ Lagrange parameters enforce the definitions of the order parameters ($\hat u$ enforces the normalization condition over the rows of the $W$ matrix). The quantity $\Xi$ is a function of the argument
\begin{equation}\label{defU}
U=\frac{2 \hat u}{2\hat s - \hat q} \ ,
\end{equation}
and is defined as
\begin{equation}\label{defxi}
\Xi (U)= \sum _{j,k\ge 2} H_j \, \Big( U \; {\bf Id}+  \boldsymbol{\cal C}\Big)^{-1}_{jk}\, H_k \qquad \text{with} \qquad H_j = 2\, \boldsymbol {\cal C} _{1j} -\phi_{0}  \ ,
\end{equation}
and ${\bf Id}$ is the identity matrix. In the above equation, the inverse is intended over the $N-1$-dimensional restriction of the matrix $U \; {\bf Id}+  \boldsymbol{\cal C}$ to entries $j,k\ge 2$.

\subsubsection{Computation of $\Xi$} 

Expanding the terms in $\Xi(U)$ in eqn. (\ref{defxi}) above, we write $\Xi(U)=\Xi_1(U)+\Xi_2(U)+\Xi_3(U)$ with
\begin{eqnarray}
    \Xi_1(U) &=&4\sum_{j,k\ge 2} \boldsymbol{\cal C}_{1j}\Big( U \; {\bf Id}+ \boldsymbol {\cal C} \Big)^{-1}_{jk}\; \boldsymbol{\cal C}_{1k}\ , \label{xi1}\\
    \Xi_2(U)&=&  \phi_0^2 \sum_{j,k\ge 2} \Big( U \; {\bf Id} + \boldsymbol {\cal C} \Big)^{-1}_{jk}\ , \label{xi2}\\
    \Xi_3(U) &=& - 4\, \phi_0 \sum_{j,k\ge 2} \boldsymbol{\cal C}_{1j}\Big( U \; {\bf Id}+ \boldsymbol {\cal C} \Big)^{-1}_{jk} \ . \label{xi3}
\end{eqnarray}

\vskip.3cm
\noindent \underline{Computation of $\Xi_1$:} Consider the $N\times N$ matrix $\boldsymbol{{\cal C}^{(N)}}$, with entries $\boldsymbol{ {\cal C}}_{ij}$ for $i,j$ comprised between 1 and $N$. Let us also define ${\bf Id^{(N)}}$ the identity matrix in dimension $N$, while ${\bf Id}$ above referred to the identity matrix in dimension $N-1$. Using block-matrix inversion formulas, we write that
\begin{equation}
   \Big(U\,{\bf Id^{(N)}}+ \boldsymbol{{\cal C}^{(N)}}\Big)^{-1}_{11}= \frac 1{\displaystyle{U+\boldsymbol{\cal C}_{11}- \sum_{j,k\ge 2} \boldsymbol{\cal C}_{1j}\Big( U \; {\bf Id}+ \boldsymbol {\cal C} \Big)^{-1}_{jk}\; \boldsymbol{\cal C}_{1k}}}
\end{equation}
The left hand side of the equation above is equal, in the large--$N$ limit, to the resolvent $g(U)$ of $\boldsymbol {\cal C}$ defined in (\ref{resolvent}). 
Using $\boldsymbol{\cal C}_{11}=\Gamma (0)=\phi_0$ and the definition of $\Xi_1(U)$, we obtain
\begin{equation}
    \Xi_1(U)= 4 \left( U + \phi_0 - \frac 1{g(U)}\right)\ .
\end{equation}

\vskip.3cm
\noindent \underline{Computation of $\Xi_2$:} Let $|v_+\rangle$ be the normalized vector with $N$ identical components, $(v_+)_i=\frac 1{\sqrt N}$. We have 
\begin{equation}
\Xi_2 (U)= N\, \phi_0^2\, \Big\langle v_+\Big| \Big( U \; {\bf Id}+  \boldsymbol{\cal C}\Big)^{-1}\Big| v_+\Big\rangle \ .
\end{equation}
For large $N$, $|v_+\rangle$ is the top eigenvector of $\boldsymbol{\cal C}$, with (extensive) eigenvalue $\lambda_+= N \int d{\bf r}\,  \Gamma({\bf r}) = N \phi_0^2$. Hence,
\begin{equation}
\Xi_2 (U)= N\, \phi_0^2\times \frac 1{U + N\phi_0^2} \to 1 \ ,
\end{equation}
in the large-$N$ limit (since $U$ remains bounded, see below).

\vskip.3cm
\noindent \underline{Computation of $\Xi_3$:} As $\boldsymbol{\cal C}_{jk}$ with $j,k\ge 2$ does not depend on the locations ${\bf r}_1^\ell$ of the place fields associated to neuron $i=1$ in the different maps $\ell$, we may substitute $\boldsymbol{\cal C}_{1j}$ in eqn.~(\ref{xi3}) with its average over those positions, equal to $\phi_0^2$. We obtain
\begin{equation}
    \Xi_3(U)=-4\, \phi_0^3\, \sum_{j,k\ge 2} \Big( U \; {\bf Id} + \boldsymbol {\cal C} \Big)^{-1}_{jk}=-4\, \phi_0\ ,
\end{equation}
in the large-$N$ limit, see calculation of $\Xi_2(U)$ above.

\vskip.3cm
\noindent \underline{Expression of $\Xi$:} Gathering the three terms above, we obtain
\begin{equation}\label{Xig}
\Xi(U)= 1 + 4\, U - \frac 4{g(U)}\ .
\end{equation}

\subsubsection{Expression of log. volume and saddle-point equations close to the critical line} 

As $\alpha$ reaches its maximal value (at fixed $\kappa$), the set of couplings satisfying the inequalities (\ref{primal}) shrink to a single solution, and we expect $s,q$ to reach the same value according to (\ref{defsq}). We therefore look for an asymptotic expression for $ \frac1{N^2} {\langle\log  Z \rangle}$ in (\ref{logvol}) when 
\begin{equation}
\epsilon = s- q \ ,
\end{equation} 
is very small and positive. In this regime, we expect the conjugated Lagrange parameters to diverge as inverse powers of $\epsilon$. More precisely, calling
\begin{equation}
\hat \epsilon = 2\hat s- \hat q \ ,
\end{equation} 
we assume that
\begin{equation}
\hat \epsilon = \frac {V}{\epsilon} \quad , \qquad \hat q = \frac{T}{\epsilon^2} \  .
\end{equation} 
as $\epsilon\to 0$.
To the leading order, we obtain 
\begin{equation}\label{logz2}
\frac1{N} {\langle\log  Z_1 \rangle} =  \frac {F(\alpha)}{2\epsilon} + O \Big( |\log \epsilon| \Big)  \ ,
\end{equation}
where $F(\alpha)$ is the extremum over $m,q,U,V,T$ of
\begin{eqnarray}\label{defga1}
F(\alpha; m,q,U,V,T) &=& V\bigg( q+U - \frac {m^2}{\Xi(U)}\bigg)   + T\bigg(1- \frac{1}V \int d\lambda \,\rho(\lambda)\frac{\lambda}{\lambda + U}\bigg)  - \alpha p (q-m^2) \int_{x}^\infty \frac{dz}{\sqrt {2\pi}} e^{-\frac{z^2}2} (z-x)^2 \quad
\end{eqnarray}
with $x=\frac{m-\kappa}{\sqrt{q-m^2}}$ and $U$ defined in (\ref{defU}). This equation is equivalent to equation (7) of the main text. Note that, in order to obtain (\ref{defga1}), the saddle point equation over $\hat m$ in (\ref{logvol}) was derived and solved for $\hat m$ explicitly. Extremizing over $U,T,V$, we obtain 
\begin{eqnarray}
V & = & \int d\lambda\,\rho(\lambda) \frac{\lambda}{\lambda + U}\ , \\
T &=& -\left( q + U - \frac{m^2}{\Xi(U)} \right) \; \int d\lambda\,\rho(\lambda) \frac{\lambda}{\lambda + U}\ , \\
%T &=& - \frac {\bigg(\displaystyle{\int d\lambda\,\rho(\lambda) \frac{\lambda}{\lambda + U}}\bigg)^2}{\displaystyle{\int d\lambda\,\rho(\lambda) \frac{\lambda}{(\lambda + U)^2}}}\times \bigg[ 1 + \frac{m^2}{\Xi(U)^2}\, \frac{d\Xi}{dU}\bigg]\ , \\
1 + \frac{m^2}{\Xi(U)^2}\, \frac{d\Xi}{dU} &=& \left( q + U - \frac{m^2}{\Xi(U)} \right) \;\; \int d\lambda\,\rho(\lambda) \frac{\lambda}{(\lambda + U)^2} \ .
%q + U - \frac{m^2}{\Xi(U)} &=& \frac {\displaystyle{\int d\lambda\,\rho(\lambda) \frac{\lambda}{\lambda + U}}}{\displaystyle{\int d\lambda \,\rho(\lambda) \frac{\lambda}{(\lambda + U)^2}}}\times \bigg[ 1 + \frac{m^2}{\Xi(U)^2}\, \frac{d\Xi}{dU}\bigg] \ .
\label{implicitU}
\end{eqnarray}
Note that the derivative of $\Xi$ with respect to $U$ can be easily computed from the derivative of $g$ with respect to $U$ according to eqn (\ref{Xig}). Following the implicit equation over $g$ in equation (9) in the main text, we find
\begin{equation}
\frac{dg}{dU}(U) = \frac 1{\displaystyle{\sum_{{\bf k}\ne {\bf 0}} \frac{\alpha\,\hat \Gamma({\bf k})}{(\alpha+ g\,\hat \Gamma({\bf k}))^2}-\frac 1{g^2}
}} \ .
\end{equation}

We may now write the saddle-point equations over $q$ and $m$, which give, after some elementary manipulation,
\begin{eqnarray}\label{implicitq}
\alpha\, p \, H(x) &=& \int d\lambda \,\rho(\lambda)\frac{\lambda}{\lambda + U} \ , \\
\frac{m}{m-\kappa} \bigg( \frac 1{\Xi(U)}-1\bigg) &=& \frac 1{\displaystyle{ \sqrt{2\pi}\,x\,e^{x^2/2}\,H(x)}}-1\ . \label{implicitm}
\end{eqnarray}
The three coupled equations (\ref{implicitU},\ref{implicitq},\ref{implicitm}) allows one, in principle, to compute $q,m,U$ and, and therefore $T,V$ and $F(\alpha)$. In addition, the optimization of $\langle \log Z\rangle$ in (\ref{logz2}) over $\epsilon$ immediately gives $F(\alpha)=0$, hence, a fourth equation to determine the critical value of $\alpha$ at fixed $\kappa$. This last equation read, after simplification according to eqn (\ref{implicitm}),
\begin{equation}\label{implicitalpha}
\frac U\kappa = m\, \bigg( \frac 1{\Xi(U)}-1\bigg) \ .
\end{equation}

\subsubsection{Large-$p$ behavior of the critical capacity} 
\label{Lp}

We now focus on the maximal capacity, obtained when $\kappa\to0$. According to (\ref{implicitalpha}), $U$ vanishes, and equations (\ref{implicitq},\ref{implicitm}) as a well as the implicit equation (9) of the main text on the resolvent $g$ give a set of two coupled equations for $x$ and the resolvent $g$:
\begin{eqnarray}
\frac 1g &=& \sum_{{\bf  k} \ne {\bf 0}} \frac{\hat \Gamma({\bf k})}{1+g\, p\, H(x)\, \hat \Gamma({\bf k})} \ ,  \label{eqrt5}\\
1-\frac 4g &=&x\sqrt{2\pi} \,H(x)\, e^{x^2/2}  \ . \label{eqrt6}
\end{eqnarray}
from which the capacity can be computed as a function of the number $p$ of points,
\begin{equation}\label{eqrt7}
\alpha_c (p)= \frac 1{p\, H(x)}\ .
\end{equation}
In practice, we can choose $x$ at will, compute $g$ from (\ref{eqrt6}), then $p$ from (\ref{eqrt5}), and, finally, $\alpha_c$ from (\ref{eqrt7}).

Remark that equation (\ref{eqrt5}) can be rewritten as 
\begin{equation} 
p\, H(x) =  G\Big(g\, p\, H(x) \Big) \qquad \text{with} \qquad
G(y) = \sum_{{\bf k} \ne {\bf 0}} \frac{y\, \hat \Gamma({\bf k})}{1+y\, \hat \Gamma({\bf k})} 
\ . 
\end{equation}
According to dimensional analysis, the large momentum scaling of the Fourier coefficients is given by
\begin{equation}
\hat \Gamma({\bf k}) \sim \frac{\phi_0^2}{\left(k \, \phi_0^{\frac 1D}\right)^{D+1}}= \frac{\phi_0^{1-\frac 1D}}{k ^{D+1}} \ ,
\end{equation}
where $k=|{\bf k}|$ and $D$ is the dimension of the physical space. We deduce that, for large arguments $y$,
\begin{equation}
G(y)\sim A_1(D) \; \phi_0 ^{\frac{D-1}{D+1}}\; y^\frac{D}{D+1}\quad \text{with}\qquad A_1(D) = \int \frac{d^D {\bf u}}{|{\bf u}|^{D+1}+1} \ .
\end{equation} 
In addition, using the asymptotic expansion of the $\erfc$ function, we have
\begin{equation}
x\sqrt{2\pi} \,H(x)\, e^{x^2/2} \simeq 1 - \frac 1{x^2} 
\end{equation}
for large $x$. Combining these expressions allows us to obtain the asymptotic relation between $x$ and $y$,
\begin{equation}
y^\frac{1}{D+1} = 4\, A_1(D)\, \phi_0 ^\frac{D-1}{D+1} \, x^2  \ . 
\end{equation}
and, to the leading order in $p$, 
\begin{equation} \label{scax}
    x\simeq \sqrt{2\log p} - \big( D+\frac 12\big) \frac{\log \log p}{\sqrt{2\log p} } \ .
\end{equation}
We then deduce the asymptotic scaling of the critical capacity given by equation (10) of the main text, with
\begin{equation}
     A(D)=\frac 1{ 8^D\, A_1(D)^{D+1} }\ .
\end{equation}

The scaling for $x$ in eqn.~(\ref{scax}) entails the following relation between the order parameters $q$ and $m$ in the large--$p$ regime,
\begin{equation}\label{qm2}
    \frac{q}{m^2} -1 \sim \frac 1{2\, \log p} \ .
\end{equation}
To interpret the consequences of the equation above, we consider a set of replicated couplings, $\{ W_{ia}\}$. For any random position $\bf r$ in map $\ell$ defining the pattern $\boldsymbol \sigma$, we define the rescaled and centered random variable
\begin{equation}
Y\big({\bf r} | \{W_{ia}\}\big) = \frac{1}{m_\ell^{a}} \bigg( \big( 2 \sigma^\ell_i -1)\, \sum _{j\ge 2} W_{ja} \, \sigma^\ell_j  - m_\ell^{a}\bigg) \ .
\end{equation}
By definition of the order parameter $m$, the average value of $Y$ vanishes:
\begin{equation}
\big\langle Y\big({\bf r} | \{W_{ia}\}\big) \big\rangle_{\bf r}  = 0\ .
\end{equation}
Equation (\ref{qm2}) implies that the variance of $Y$ is
\begin{equation}
\big\langle Y\big({\bf r} | \{W_{ia}\}\big) ^2 \big\rangle_{\bf r}  \simeq \frac 1{2\log p}\ ,
\end{equation}
as $p$ gets large and the load takes its maximal value (critical capacity).
In other words, the standard deviation of $Y$ scales as $(\log p)^{-1/2}$ for large $p$. We thus expect that the $k^{th}$ cumulant of $Y$ will scale as $(\log p)^{-k/2}$. Under this assumption, the distribution of the stability $t$ has mean value $m$ and fluctuations of the order of $\Delta t= m/\sqrt{\log p}$. These fluctuations are negligible in the large--$p$ limit, since  resolution of the saddle-point equation (\ref{implicitU}) shows that \begin{equation}
    m\simeq  \frac D4 - \frac{D^2}{256\, (\log p)^3} +o \left( \frac 1{(\log p)^3} \right)
\end{equation}
at the critical point. Hence, $\Delta t \sim (\log p)^{-1/2}$ is smaller and smaller as $p$ increases, and the distribution of $t$ is well approximated by a Gaussian in the large--$p$ limit. The  Gaussian approximation obtained by discarding all powers of $\hat t$ of order $\ge 3$ in eqn (\ref{expansion_hatt}) in our quenched PF theory is therefore expected to be exact in this limit.

\subsection{Comparison between Quenched PF theory and SVM}

\subsubsection{Values of couplings}

\begin{figure}[ht]
	\centering
	\includegraphics[width=.483\textwidth]{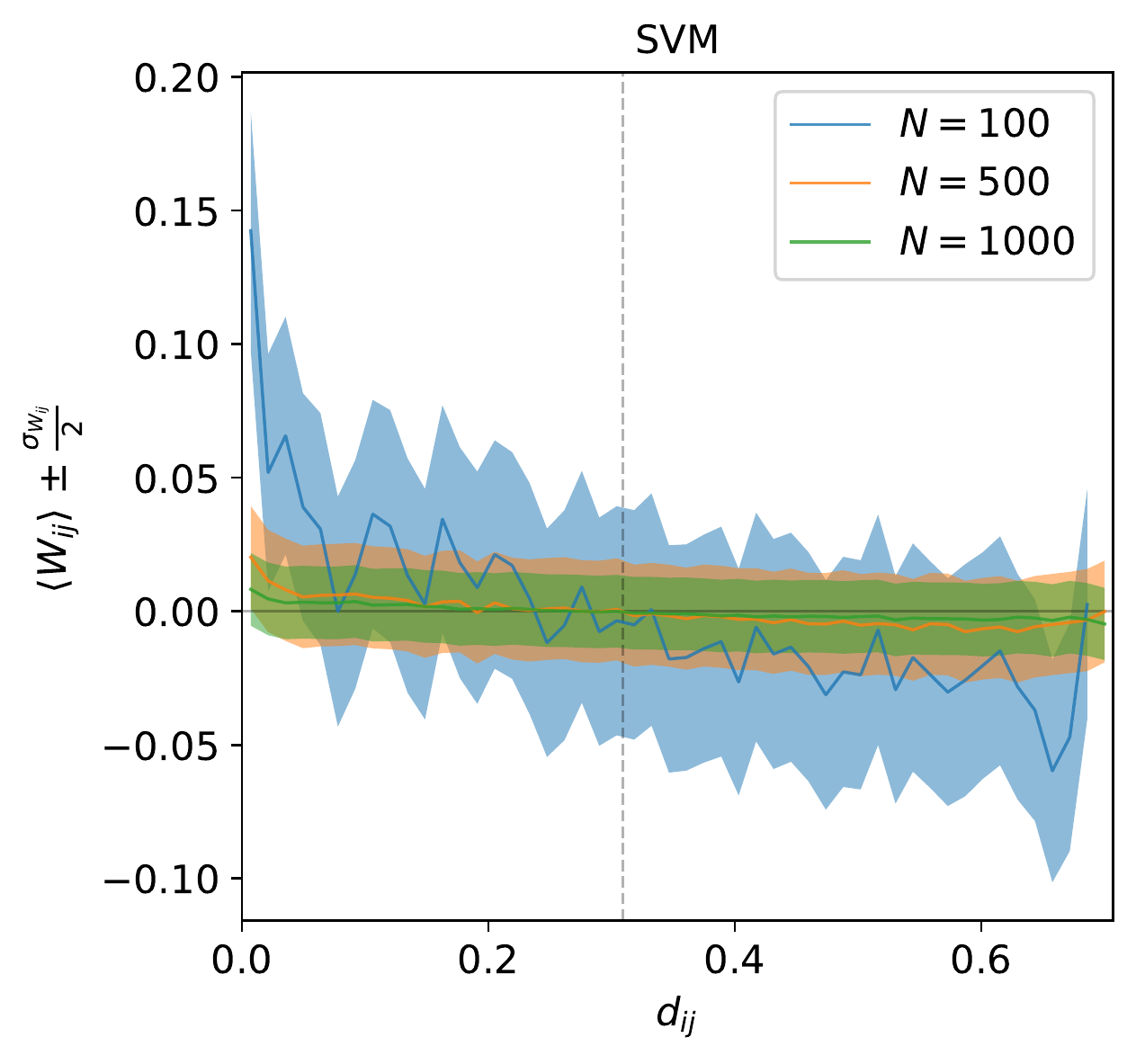}
	\includegraphics[width=.483\textwidth]{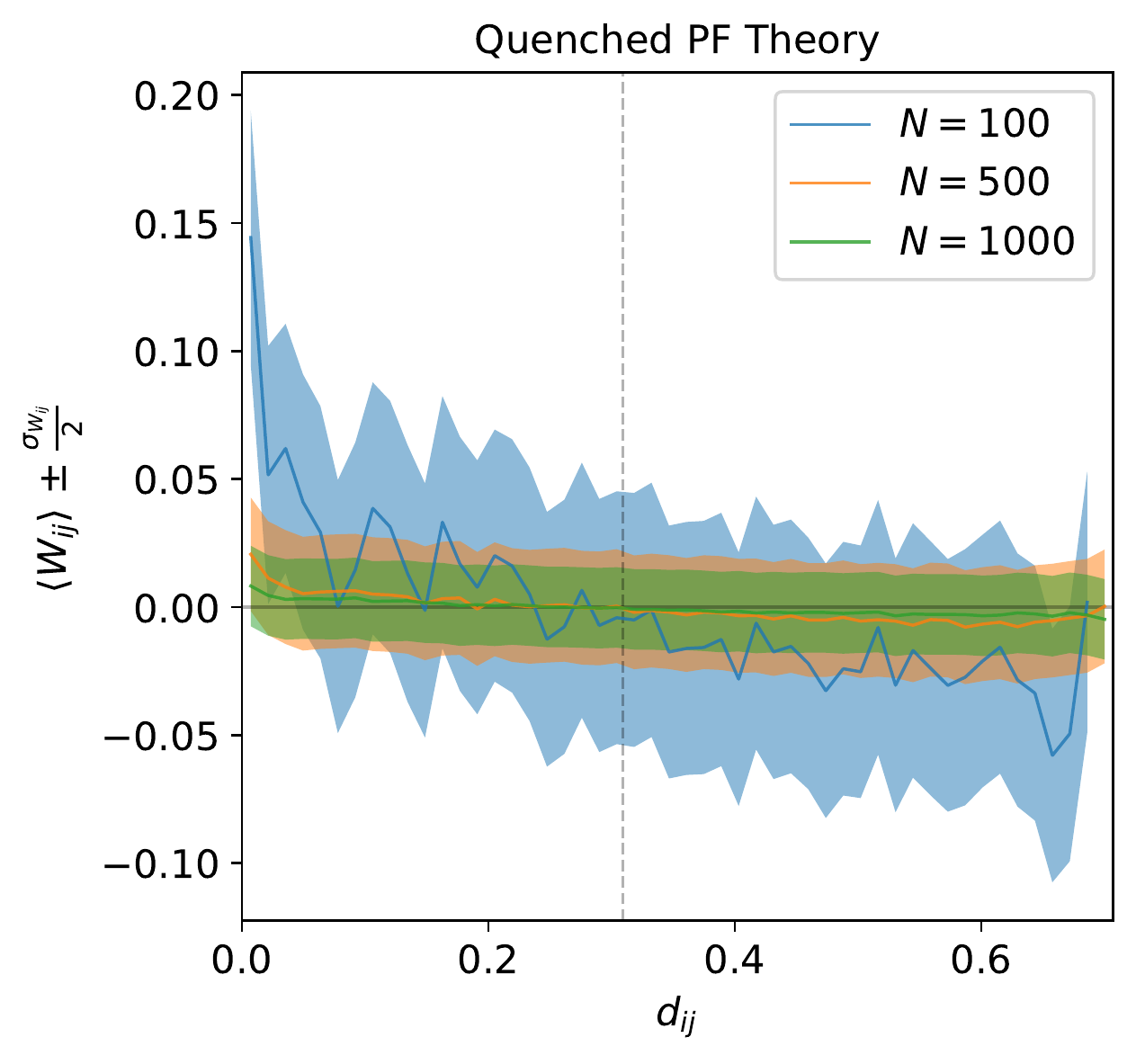}
	\includegraphics[width=.483\textwidth]{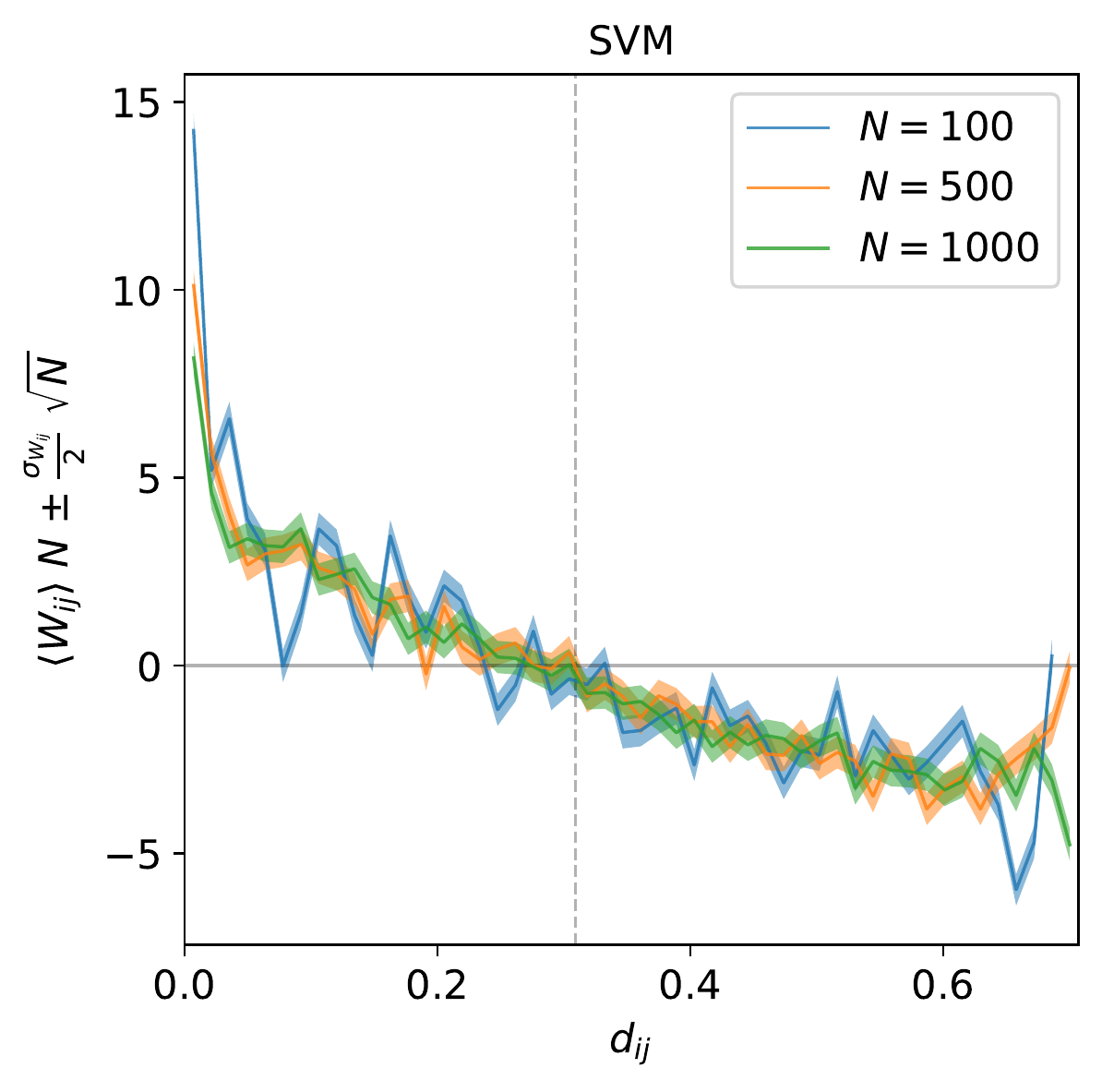}
	\includegraphics[width=.483\textwidth]{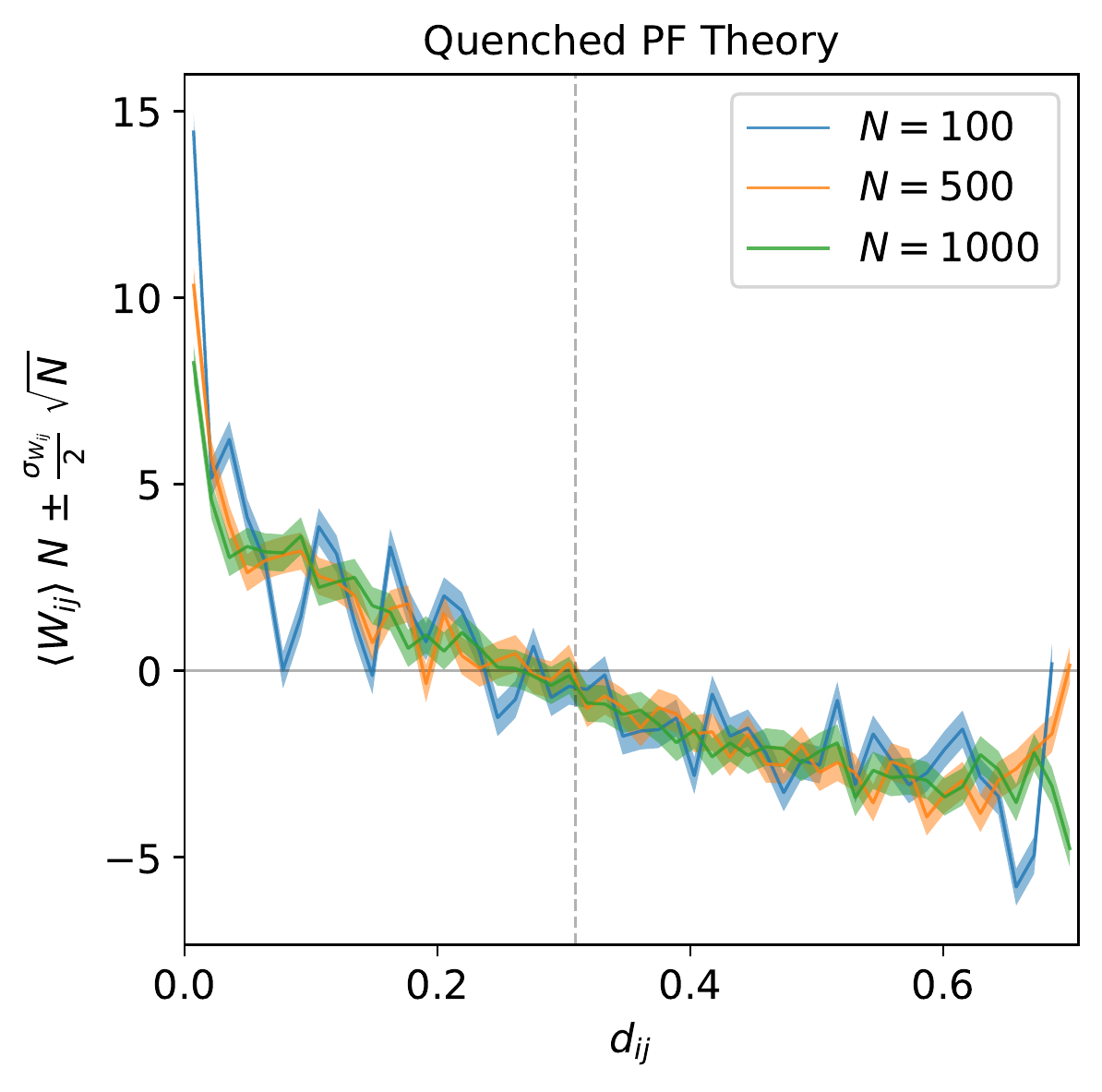}
	\caption{Comparison of couplings obtained with SVM (left) and with the Quenched PF Theory (right). Top: Dependence of couplings on $N$. 
	Bottom: Dependence of the couplings on distance; The vertical line locates the radius $r_c$ of the PF. These results were obtained for $D=2$, $\phi_0=.3$, $\alpha=.1$, $p=5$; we have averaged over $100$ different realizations of the $p$ positions at fixed PF centers for the SVM results. Space was divided in $50$ bins with values ranging from $0$ to $\sqrt{2}/2$ (the maximal distance achievable in unit square with periodic boundary conditions). Couplings were then put in the corresponding bins for all maps, and the averages and standard deviations were plotted as  functions of the bin centers. Average couplings and associated standard deviations with quenched PF theory were computed with $(23)$ and $(24)$ of \cite{monasson}, with the substitution $\alpha_c \to p \, \alpha_c$ as the number of patterns is here $p \times L$.}
        \label{fig:FIG_IIE}
\end{figure}

Figure~\ref{fig:FIG_IIE} compares how the couplings $W_{ij}$ depend on the size of the network, $N$, and of the distances between the PF of neurons $i,j$ in the maps. We generally find that the couplings $W_{ij}$ obtained by SVM and the `thermal' averages $[W_{ij}]$ predicted by the quenched PF theory for a fixed set of PF centers are in excellent agreement, see equations $(23)$ and $(24)$ in \cite{monasson} for details on the calculation of the average couplings and associated standard deviations.
\begin{itemize}
    \item Both sets of couplings have mean values scaling as $\frac{1}{N}$ and standard deviations scaling as $\frac{1}{\sqrt{N}}$.
    \item  The sign of couplings depend on the distance between their PF centers in the maps. We get excitatory couplings for distances up to the radius of the PFs ($r_c=\sqrt{\frac{\phi_0}{\pi}}$ in $D=2$), and inhibitory interactions for larger distances. 
\end{itemize}

\subsubsection{Dependence on $\phi_0$}

\begin{figure}[ht]
	\centering
	\includegraphics[width=.5\textwidth]{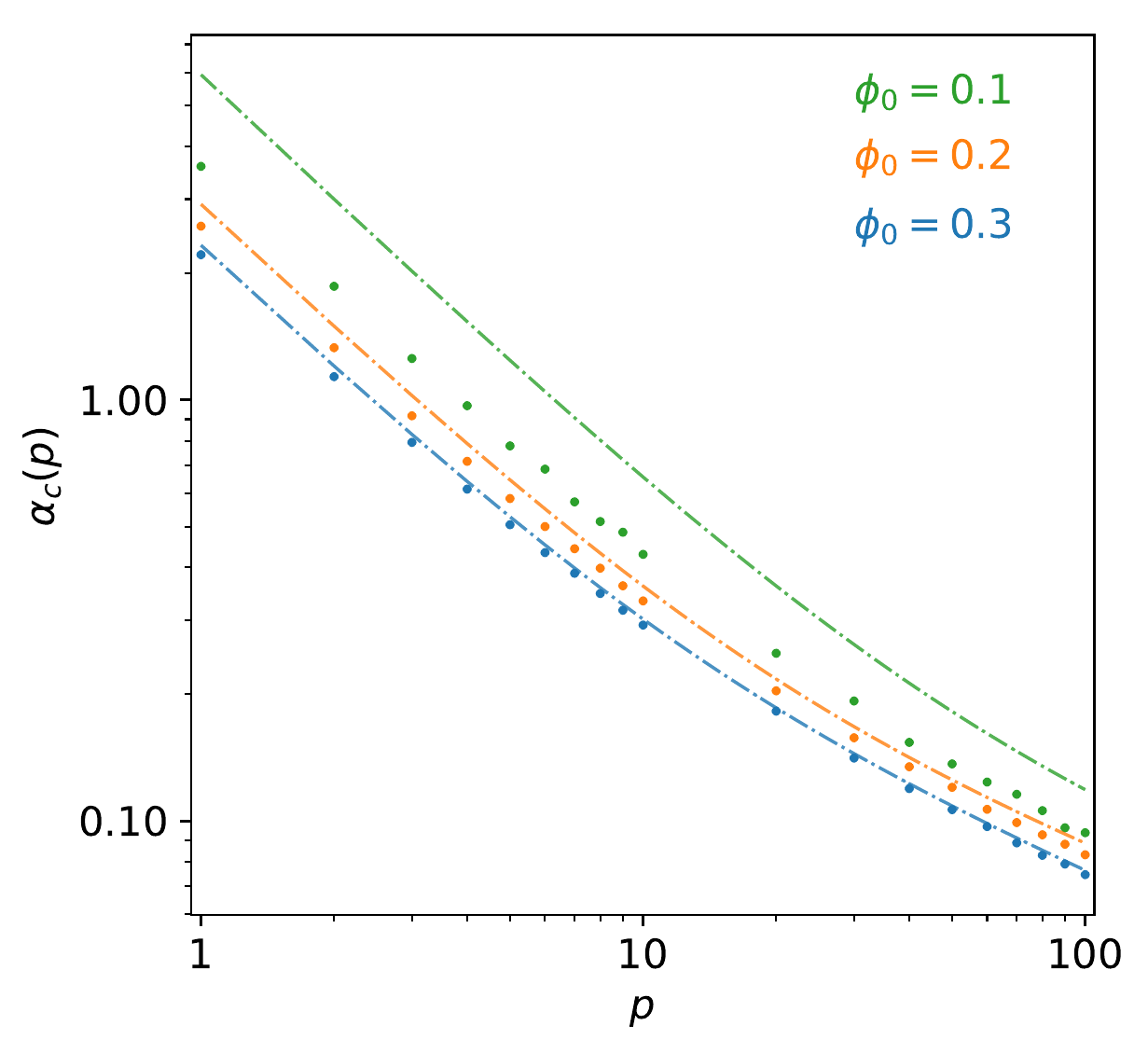}
	\caption{Scaling cross-over of $\alpha_{c}(p)$ vs. $p$ for different values of $\phi_0$. Quenched PF Theory (dashed-dotted lines) gets closer to SVM (scatter plots) as $p$ increase, the value of $p$ for which Quenched PF Theory and SVM matches increase as $\phi_0$ decrease.  We use for this results $D=2$, $N=5000$, and we  have averaged over $50$ different realization of the environments and different realizations of the $p$ positions.}
        \label{fig:FIG_IID}
\end{figure}

In Fig.~\ref{fig:FIG_IID} we show that the value of $p$ such that the results obtained with the quenched PF theory and with SVMs match increases as $\phi_0$ decrease.

An analysis of equations (\ref{eqrt5},\ref{eqrt6}), valid in the small $\phi_0$ limit, indicate that this minimal value of $p$ scales as
\begin{equation}
    p_{match}(\phi_0) \sim \frac{e^{ 1/(8\phi_0)}}{\phi_0^{3/2}} \ ,
\end{equation}
and becomes very large as $\phi_0$ is small. Realistic values for $\phi_0$ are reported in the experimental literature \cite{mizuseki} and \cite{hussaini} to range between .2 and .3.